\PassOptionsToPackage{table}{xcolor}
\documentclass[12pt, table]{article}
\usepackage{geometry}
\usepackage{authblk}
\usepackage{hyperref}
\usepackage{appendix}
\usepackage{graphicx, wrapfig}
\usepackage{amssymb}
\usepackage{amsmath}
\usepackage{amsthm}
\usepackage{esvect}
\usepackage{epigraph}
\usepackage{fancyhdr}
\usepackage{xcolor}
\usepackage{caption, subcaption}
\usepackage{geometry}
\usepackage{rotating}
\usepackage[table]{xcolor}
\usepackage{booktabs}
\usepackage{color,soul}
\newtheorem{definition}{Definition}

\title{\Large \textbf{Impacts of Economic Policies on Wealth Distribution in Token Economies}\\ \large How do endogenous and exogenous economic policies impact wealth distribution in token economies?}
\author{R. Sadykhov, Dr.G. Goodell and Prof.P. Treleaven}
\affil{University College London}
\date{January 2026}

\begin{document}

\begin{titlepage}
\maketitle
\thispagestyle{empty}
\begin{abstract}
In this paper, we analyse the impacts of exogenous and endogenous factors on wealth distribution in the Bitcoin token economy, where wealth distribution refers to the distribution of BTC between economic participants or groups of economic participants. The objective of the paper is to analyse the impact of economic policies on wealth distribution in the Bitcoin ecosystem.\par
Different macroeconomic and microeconomic time series are used to eliminate noise in the wealth distribution time series, and the causality analysis is performed between Bitcoin Improvement Proposals (i.e., BIPs) and the cleaned wealth distribution data to reveal possible patterns in the impacts that the endogenous policies have on wealth distribution in token economies. Lastly, a structure for economic policy taxonomy in token economies is proposed where different the policy implementations are illustrated by existing BIPs.\par
This approach highlights the actions available to the policy makers, as well as providing a technique for analysis of policy impacts in token economies and their categorization.
\end{abstract}
\end{titlepage}

\section{Introduction}
Since the introduction of blockchain in 2008, token economies (as defined by Sadykhov et al. \cite{DeTEcT}) have become some of the most popular financial systems, and as their growth continues many wonder how can such decentralized economic mechanisms be regulated and what policies can financial regulators introduce to ensure that token economies are stable, comply with anti-money laundering and anti-terrorist funding schemes, and are taxable.\par
In order to understand what policies must be set, it is important first to evaluate what policies have an impact on token economies. The level at which the policies are applied should also be considered: would setting an administered interest rate for overnight repurchase agreements help to incentivize economic activity in a token economy, or does it require a specific incentive program to stimulate consumption? From this question, it is evident that there are different types of policies that can target a token economy and we would be referring to these as exogenous and endogenous policies.\par
Exogenous policies are the policies that are introduced outside of the token economy's scope, and these policies are usually directed at the ``global'' state of an economy (e.g., setting a tax bounds for different income tax categories, issuing capital gains tax credit, etc.). In contrast, endogenous policies are the policies set inside of the token economy (e.g., maximum supply) and are directed purely at the economic activity inside it. Note, that we do not claim that every exogenous policy is unrellated to the state of a token economy, there can be policies that introduce a tax on realized income which is specific to a token economy, and vice versa, there can be endogenous policies whose impact spills over into the global economy.\par
From these definitions of exogenous and endogenous policies, we deduce that exogenous policies tend to have a much larger scope, and therefore, in order to study the effect of endogenous policies we somehow have to filter out the ``noise'' produced by the exogenous policies and by non-policy related events. Without this step we will not be able to determine whether the endogenous policy had an impact, was it implemented in response to something that has affected the token economy, or did not have any impact at all.\par
Finally, to conduct our research, we measure the policy impact using wealth distribution (i.e., distribution of wealth between buckets of Bitcoin addresses). The reason for this choice is because wealth distribution demonstrates the economic activity in an economy as economic activity is the redistribution of wealth in an economy. Also, wealth distribution is a great benchmark to see how policies impact different compartments of the population, as very often, policy makers target different agent categories \cite{DeTEcT} based on their wealth or income level.\par
With all of the above points, the aim of this paper is to examine how different exogenous and endogenous policies impact the wealth distribution in a token economy, and to develop a method for performing such analysis for a generic token economy. We also aim to map endogenous policies to their exogenous counterparts in the economic policy taxonomy.

\section{Scope}
\label{section:Scope}
In this paper we will perform analysis described above using the data from the Bitcoin token economy. The reason for this choice is the availability of high quality data on Bitcoin, and in particular the existence of Bitcoin Improvement Proposals \cite{BIPs} (i.e., BIPs). BIPs are organised into a repository of proposals submitted by Bitcoin community to improve the design of Bitcoin. These improvement proposals range from cybersecurity and performance improvements, to the design of Bitcoin's economic mechanism. Not all BIPs will be relevant to the construction of economic policy taxonomy for token economies, but we still will use them in our analysis of wealth redistribution. Therefore, for all purposes we will be treating BIPs as the endogenous policies of the Bitcoin token economy.\par
The Bitcoin wealth distribution data that we use, along with the Bitcoin close price data, are taken from CCData \cite{CCData}. The wealth distribution data is broken down into 10 buckets each of which is a sum of holdings of all addresses that satisfy that buckets' bounds. For example, the bucket labelled \emph{From 1 to 10} is a time series of a sum of all Bitcoin holdings that belong to addresses who hold between 1 and 10 Bitcoins at the given time.\par
First, note that the sum of all 10 buckets is nothing else but the time series of Bitcoin's current supply. Second, the addresses in this buckets are not constrained to stay in the same bucket for the duration of the time series, and therefore, the number of addresses in each bucket can change in time just as well as the wealth of that bucket.\par
The software that we will use to perform the analysis is written in combination of Rust and Python programming languages, and the libraries used to analyse the data are DigiFi \cite{DigiFi}, Statsmodels \cite{StatsModels}, SciPy \cite{SciPy} and Numpy \cite{Numpy}.\par
To perform the rest of our analysis, we need exogenous policies time series, macroeconomic and microeconomic data. The data in these three categories we see as the potential variables that may have a significant impact on wealth distribution, and therefore, we would like to check whether there is any noise accossiated with the time series that represent these categories. The time series used for the analysis are listed below.\par
\subsection{Exogenous Policies Data}
The exogenous policies are the policies that are set outside of scope of a token economy. To represent exogenous policies we use the time series defined below. Note that in brackets we place the source, time series identifier used by the source, measurement frequency, and the last date in the time series.\par
\begin{itemize}
	\item Federal Funds Effective Rate (FRED \cite{FRED}, DFF, Daily, 28-08-2025)
	\item Personal Current Taxes (FRED \cite{FRED}, W055RC1, Monthly, 01-07-2025)
	\item 5-Year High Quality Market (HQM) Corporate Bond Par Yield (FRED \cite{FRED}, HQMCB5YRP, Monthly, 01-07-2025)
    \item 10-Year High Quality Market (HQM) Corporate Bond Par Yield (FRED \cite{FRED}, HQMCB10YRP, Monthly, 01-07-2025)
    \item 30-Year High Quality Market (HQM) Corporate Bond Par Yield (FRED \cite{FRED}, HQMCB30YRP, Monthly, 01-07-2025)
    \item 5-Year High Quality Market (HQM) Corporate Bond Spot Rate (FRED \cite{FRED}, HQMCB5YR, Monthy, 01-07-2025)
    \item 10-Year High Quality Market (HQM) Corporate Bond Spot Rate (FRED \cite{FRED}, HQMCB10YR, Monthly, 01-07-2025)
    \item 30-Year High Quality Market (HQM) Corporate Bond Spot Rate (FRED \cite{FRED}, HQMCB30YR, Monthly, 01-07-2025)
	\item British Government Securities, Nominal Par Yields, 5 year (IADB \cite{IADB}, IUDSNPY, Daily, 29-08-2025)
	\item British Government Securities, Nominal Par Yields, 10 year (IADB \cite{IADB}, IUDMNPY, Daily, 29-08-2025)
	\item British Government Securities, Nominal Par Yields, 20 year (IADB \cite{IADB}, IUDLNPY, Daily, 29-08-2025)
	\item British Government Securities, Nominal Zero Coupon Yields, 5 year (IADB \cite{IADB}, IUDSNZC, Daily, 29-08-2025)
    \item British Government Securities, Nominal Zero Coupon Yields, 10 year (IADB \cite{IADB}, IUDMNZC, Daily, 29-08-2025)
    \item British Government Securities, Nominal Zero Coupon Yields, 20 year (IADB \cite{IADB}, IUDLNZC, Daily, 29-08-2025)
\end{itemize}
The reason we selected these features as the exogenous policy data is beacuse authorities and regulators can directly immpact them, and through these, impact the rest of the economy. This specific selection was also made because the data for these time series was accessible and all of the readings in these time series span at least from 2008 to August 2025, which is important as it maps to the range of the data for our Bitcoin supply distribution data. There are plenty of other time series that could be considered here, but we will limit our analysis to this data only.\par
\subsection{Macroeconomic Data}
Below are the time series that we use to explain some of the noise that exists in the wealth distribution time series due to macroeconomic factos. Just like in the case with exogenous policies, there are some time series that may better explain some of the variance of the wealth distribution buckets, but we stick to these as they are accessible and cover a lot of diverse macroeconomic factors.\par
\begin{itemize}
	\item Total Nonfarm All Employees (FRED \cite{FRED}, PAYEMS, Monthly, 01-07-2025)
	\item Unemployment Rate (FRED \cite{FRED}, UNRATE, Monthly, 01-07-2025)
	\item M1 (FRED \cite{FRED}, WM1NS, Weekly, 04-08-2025)
	\item M2 (FRED \cite{FRED}, WM2NS, Weekly, 04-08-2025)
	\item Market Value of Gross Federal Debt (FRED \cite{FRED}, MVGFD027MNFRBDAL, Monthly, 01-07-2025)
    \item Market Value of Privately Held Gross Federal Debt (FRED \cite{FRED}, MVPHGFD027MNFRBDAL, Monthly, 01-07-2025)
	\item Consumer Price Index for All Urban Consumers: All Items in U.S. City Average (FRED \cite{FRED}, CPIAUCSL, Monthly, 01-07-2025)
	\item Gold Price Against US Dollar (Yahoo! Finance \cite{YahooFinance}, GC=F, Daily, 26-08-2025)
\end{itemize}
\subsection{Microeconomic Data}
Lastly, we selected a microeconomic time series that we believe is also relevant to the analysis, as semiconductors are used by the graphics cards, which are often tied to the Bitcoin mining, so we decided to include this time series to check whether it impacts the wealth distribution data.\par
\begin{itemize}
	\item Producer Price Index by Industry: Semiconductor and Other Electronic Component Manufacturing (FRED, \cite{FRED}, PCU33443344, Monthly, 01-07-2025)
\end{itemize}

\section{Impact of Exogenous Factors on Wealth Distribution}
\label{section:ImpactOfexogenousFactorsOnWealthDistribution}
In this section we examine the significance of different features described in Section \ref{section:Scope} for Bitcoin's wealth distribution, how do they impact the wealth distribution buckets, and consequently, we clear the wealth distribution buckets from the noise introduced by these features to facilitate further analysis of the cleared data. We start by creating a model for each of the wealth distribution buckets in terms of the features from Section \ref{section:Scope}. Once we filter out the insignificant features from the model, we use this model to extract the noise term $\epsilon$ that the features fail to explain, as we consider it to be the the data that is cleaned from the impact of exogenous factors, and so we use it to analyze the impacts of BIPs in the next section.\par
\subsection{Data Pre-Processing}
\label{subsection:DataPreProcessing}
Before proceeding with our analysis, we must transform all of the time series so that they become stationary. The motivation for this is that most economic and financial time series are non-stationary and they usually experience some drift over time (e.g., exponential-like rise in \emph{Personal Current Taxes} from 1959 to 2025). This property of time series can skew our analysis as if we pick a large enough time frame and two time series that are non-stationary, there will always be a relationship between them (e.g., positive Pearson correlation) since both time series drift away from their starting point over time.\par
To mitigate this phenomenon we must apply data transformations to the time series in order to convert these time series to stationary time series. For that we use two distinct transformations: first-order differencing and log change transformation.\par
First-order differencing is a transformation where for every time step in a time series we subtract previous value in the time series from the current value in the time series so that
\begin{equation}
	\Delta x_{t} = x_{t} - x_{t-1}
\end{equation}
where $\Delta x_{t}$ is the transformed value of the series. Note that this transformation produces a time series which is one element shorter than the original time series since the initial value of the series does not have a value before it in the time series. This transformation we apply to the following time series:
\begin{itemize}
	\item Federal Funds Rate
	\item 5 Year HQM Corporate Bond Par Yield
	\item 10 Year HQM Corporate Bond Par Yield
	\item 30 Year HQM Corporate Bond Par Yield
	\item 5 Year HQM Corporate Bond Spot Rate
	\item 10 Year HQM Corporate Bond Spot Rate
	\item 30 Year HQM Corporate Bond Spot Rate
	\item 5 Year Gilts (Nominal Par Yield)
	\item 10 Year Gilts (Nominal Par Yield)
	\item 20 Year Gilts (Nominal Par Yield)
	\item 5 Year Gilts (Nominal Zero Coupon Yield)
	\item 10 Year Gilts (Nominal Zero Coupon Yield)
	\item 20 Year Gilts (Nominal Zero Coupon Yield)
	\item Unemployment Rate
	\item Gold Price Against USD
\end{itemize}
All of the time series listed above are quoted as a rate, so by applying the first-order differencing we obtain time series of changes in the original time series. The reason we want a time series of changes in the original is because we want a stationary time series that tells us how much the original one grew or declined by independently from the previous time steps, so that we can compare the growth/decline patterns across different time series. Under this transformation the distribution of increments $\Delta x_{t}$ is not skewed due to their size since a rise and decline by fixed $\Delta x_{t}$ leaves the original value $x_{t}$ unchanged (Conversely, assume a case where $x_{t}$ is increased by 5\% and then decresed by 5\% - the resulting value wil not be equal to $x_{t}$, and so the distribution of increments might be skewed towards positive increments $\Delta x_{t}$ under a simple percent change transformation).\par
The second transformation type we use is the log-change transformation, which is defined by
\begin{equation}
	R_{t} = ln(x_{t}) - ln(x_{t-1})
\end{equation}
where $R_{t}$ is the log-change of the original time series. As before, this time transformation produces a time series which is one element shorter than the original one. We apply this transformation to all time series excluding the ones listed above under the first-order differencing, and we also apply log-change to wealth distribution time series themselves. As before, log-change transformation yields time series that are stationary and it also eliminates the skewness bias described above.\par
Lastly, by applying these transformations to original time series we obtained their stationary counterparts, we now run an Engle-Granger two-step cointegration test with a $5\%$ confidence level between all of the transformed independent variables (i.e., our features' time series) and all of the transformed dependent variables (i.e., wealth distribution buckets) to confirm that there is a long-term stable relationship between the features and wealth distribution time series. The cointegration test was passed by every pair of independent and dependent variables.\par
\subsection{Linear Regression Analysis of Wealth Distribution}
Having transformed the data into stationary time series, we now attempt to explain away some of the variance in the wealth distribution that potentially belongs to external factors. To do so we propose to design a linear regression model for every wealth distribution bucket in terms of the exogenous factor time series we have at our disposal.\par
The reason for using linear regression is that it produces a good estimate for the linear dependency between independent and dependent variables, and since many relationships in econommics and finance are considered to be linear, linear regression is a fitting tool to use. In section \ref{subsection:DataPreProcessing} we have also ensured that the time series we use are stationary and the linear combination of stationary time series is also stationary, which means that we can use linear regression to model the change in wealth distribution using the independent variables we have proposed.\par
Linear regression also comes with a useful toolkit for systematically filtering out insignificant independent variables by performing a t-test on the linear regression slope and a check for colinearity between independent variables by computing the variance inflation factor. However, before applying those, we must construct a base model with all available time series. In the context of this paper we refer to such models as the ``global linear regression'' models as we regress over all of the available data. A global linear regression is of the form
\begin{equation}
\label{equation:GlobalLinearRegression}
	Y = \alpha + \beta_{1}X_{1} + \beta_{2}X_{2} + ... + \beta_{n}X_{n} + \epsilon
\end{equation}
where $Y$ is the transformed wealth distribution bucket and $X_{i}$'s are the transformed independent variables. Note that since we have 10 wealth distribution buckets we will have 10 different global linear regression models.\par
Another caveat of this approach to explaining the variance in wealth distribution is that the time series we use as the features $X_{i}$ of the model all have different lengths and different frequencies. This means that we had a choice of whether to do an inner join on the data or perform a left join of features with the respective wealth distribution bucket and fill-in the missing data. Since we wanted to preserve as much of the relationship as possible, we decided to downsample our data set and use monthly readings for all variables, which resulted in the data set with 140 historical data points for each of the 10 models. We consider this to be sufficient sample size to run linear regression on in order to establish the significance of the independent variables.\par
After running all 10 global linear regression models, we have obtained the estimators $\alpha$ and $\beta_{i}$ for the model along side their standard errors $SE_{\alpha}$ and $SE_{\beta_{i}}$, which are used in the t-tests. For the t-tests, we assumed that the null hypothesis is $\beta_{i}=0$ (i.e., there is no relationship between the independent variable $i$ and the wealth distribution bucket) since we have never fitted the model before and we want to use the global linear regression to check for the existence of relationships between our independent and dependent variables. The results of the t-tests for all global linear regression models are presented in Appendix \ref{appendix:GlobalLinearRegressionPValues}.\par
Each column name in these tables is a wealth distribution bucket, and once transformed, represents $Y$ from equation \ref{equation:GlobalLinearRegression}, while every row name is the name of the feature before it was transformaed to become $X_{i}$. The way these tables can be interprited is that every column represents a model for the specific wealth distribution bucket where the row names are the features we used in this model to predict the changes in this wealth distribution bucket. The tables' cells show p-values of the linear regression slope t-test.\par
To check for significance we use a $5\%$ confidence level - the highlighted cells demonstrate the significant relationships in the independent variables and wealth distribution buckets. However, we must now filter out all of the insignificant features for every model, run the linear regression and again check for significance of each feature again. Once we have done that, we discover that some independent variables that were previously significant became insignificant. Thus we carry on with this cycle until only the variables that are significant according to the t-test at $5\%$ confidence level remain in our models.\par
The results for completely filtered models are presented in Appendix \ref{appendix:FilteredLinearRegressionPValues}. From these tables we see that the only significant independent variables left are \emph{Federal Funds Rate} and \emph{M2 (US)} that contribute to \emph{From 0 to 0.001} and \emph{From 10 to 100} wealth distribution bucket respectively.\par
\subsection{Cleaning of Wealth Distribution Data}
Having created the models for the wealth distribution buckets, we now use these models to find the error term $\epsilon$ that we can use for our analysis of BIPs and their impacts. Notice that $\epsilon$ for all wealth distribution buckets except for \emph{From 0 to 0.001} and \emph{From 10 to 100} will be simply the transformed wealth distribution buckets, since they don't have any features in their model to explain the variance.\par
The only models with significant parameters according to the t-tests are:
\begin{equation}
	Y_{\text{From 0 to 0.001}} = 0.02188 X_{\text{Federal Funds Rate}} + \epsilon
\end{equation}
and
\begin{equation}
	Y_{\text{From 10 to 100}} = 0.12899 X_{\text{M2 (US)}} + \epsilon,
\end{equation}
where we included the slope coefficients for the models, and the p-values for $X_{\text{Federal Funds Rate}}$ and $X_{\text{M2 (US)}}$ are $0.000010$ and $0.001555$ respectively. Note that both models do not have an intercept as the itercepts in either model did not have a significant t-score.\par
Since we established the shapes of the models, we can now obtain the error terms for these models, which we assume to be cleaned data that is not explained away by the exogenous factors. We will see that this assumption is important in the next section when we will by studying the impact of BIPs on this cleaned data.\par
\subsection{Summary}
In this section we have described a procedure by which we explain away the noise in wealth distribution buckets attributed to the exogenous factors to the Bitcoin token economy. We have transformed the data into stationary time series to eliminate undesired side effects of processing raw data. Then we constructed multivariate linear regression models which we used to remove the insignificant independent variables from our analysis, and then to clean the wealth distribution data to obtain the data the variance in which we assume is primarily explained by the Bitcoin Improvement Proposals.

\section{Impact of Endogenous Policies on Wealth Distribution}
\label{section:ImpactOfEndogenousPoliciesOnWealthDistribtuion}
In this section we check what Bitcoin Improvement Proposals (BIPs) can be considered as financial policies, examine their possible causal relationship with the Bitcoin wealth distribution buckets, and compare the results from different causality testing techniques. To ensure that the wealth distribution is not impacted by any exogenous factors, we have filtered out the noise produced by these such factors in the previous section. We refer to the supply distribution where the imapct of exogenous factors has been removed as the cleaned wealth distribution.\par
The analysis procedure employed here can be broken down into the following steps:
\begin{enumerate}
	\item Select a set of BIPs to perform an analysis with
	\item Create a signal time series out of the selected subset of BIPs
	\item Run a causal relationship test between the BIPs signal time series and the cleaned wealth distribution
\end{enumerate}
Following this process, we can determine whether a subset of BIPs potentially causes the changes in the wealth distribution. However, there are two major questions that arise from the process described above: how do we select a subset of BIPs to test whether they cause the changes in the wealth distribution; and what causation test do we use to test for the existence of the relationship. We attempt to answer these questions in the two subsequent sections respectively.\par
\subsection{Filtering of BIPs Data}
In an ideal world, we would have a label on each BIP that represents whether this BIP is an economic policy or not, which then we would be able to use in our analysis. However, even if that were true, there would still be a number of ``non-economical'' BIPs that would probably cause some wealth redistribution that would must be included in the analysis. There is also no possibility to run the causation tests iteratively creating samples with different combinations of BIPs as at the moment of writing there are 176 BIPs whcih makes an iterative sampling technique highly impractical.\par
Due to these issues, we decided on a different approach for selecting the  ``economic policy'' BIPs and running the causality tests. Instead of having one subset of BIPs, we decided on having the following four subsets:
\begin{itemize}
	\item \textbf{All BIPs}: A set that contains all available BIPs
	\item \textbf{All Economy-Related BIPs}: BIPs that we manually selected that seem to be economic policies - more on the selection process later
	\item \textbf{Major Economy-Related BIPs}: BIPs that are considered to have large impact according to different sources
	\item \textbf{All Economy-Related BIPs Except the Major Economy-Related Ones}
\end{itemize}
In the further sections we elaborate on our choices and methodology for selecting BIPs for these subsets.\par
\subsubsection{All BIPs}
The fist category (i.e., \emph{All BIPs}) is a set of BIPs that will be used as the benchmark data for the control test, and since most of the BIPs are distributted across the time span of the cleaned wealth distribution data, we expect the signal time series containing all BIPs to have too much noice to pass most causality tests available.\par
\subsubsection{All Economy-Related BIPs}
\label{subsubsection:AllEconomyRelatedBIPs}
The second category (i.e., \emph{All Economy-Related BIPs}) represents BIPs that we consider to be economic policies or BIPs that we believe have the potential to cause wealth redistribution. One of the objectives of this paper is to provide an initial framework for categorizing policies in token economies, but due to nonexistence of such yet we had to go over all the BIPs and manually select the BIPs that we think broadly fit under the following categories:
\begin{itemize}
	\item Improvements Aimed Directly at Supply or Liqudity: 9, 42, 100, 101, 102, 103, 104, 105, 106, 107, 109, 152, 331
	\item Transaction Improvements or Upgrades: 11, 13, 16, 65, 66, 68, 78, 79, 112, 115, 116, 118, 125, 127, 129, 134, 146, 174, 322, 324, 330, 352, 370, 373
	\item Embedded Financial Instruments and Solutions: 197, 199, 300, 345
	\item Security and Privacy Improvements or Failures: 30, 50, 53, 54, 151, 351
	\item Hierarchical Deterministic Wallets: 32, 39, 43, 44, 49, 84, 88, 175
	\item SegWit: 91, 141, 148, 173
	\item Taproot: 326, 327, 341, 343, 371
	\item Improvements that Simplify or Popularize the Bitcoin Economy: 1, 21, 47, 61, 70, 72, 75, 111, 380
\end{itemize}
Note two things: first, this categorization for structuring economic policies in a token economy is only an initial suggestion for the framework and is not necessarily exaustive; second, there are BIPs that can satisfy multiple categories (e.g., \emph{BIP 39} proposes an introduction of mnemonic code for generation of HD wallets, which can be categorized under the HD wallets improvement or as the improvement that simplifies/popularizes Bitcoin to its end users), in which case we place it under a category that has a more limiting definition (e.g., \emph{BIP 39} goes into the HD wallets category).\par
By categorizing BIPs into these categories it helps us to understand what part of the token economy they are targeting and will help us in building the economic policy taxonomy for token economies in Section \ref{section:PolicyTaxonomyOfTokenEconomies}. The reason we do not use these categories for our analysis directly is because they have been manually selected and as explained above some of the BIPs have stong overlap with other categories so we would have to come up with a mechanism for deciding whether to include these BIPs or not, which will make the results even more subjective.\par
Lastly, notice that we have purposefully defined \emph{Hierarchical Deterministic Wallets}, \emph{SegWit} and \emph{Taproot} as separate categories. The motivation behind this is that these were major series of updates to many different parts of Bitcoin that have caused a lot of discussions in the Bitcoin community at the time of their releases. Therefore, we hypothesise that these events would have caused major changes in the wealth distribution and since each of them touched upon a lot of different parts of the Bitcoin ecosystem, we decided to treat BIPs related to them as separate categories.\par
\subsubsection{Major Economy-Related BIPs}
As mentioned in the previous section, there are some BIPs that stand out amongst the other ones as they proposed or described major changes to the Bitcoin token economy. We have counted five such BIPs, notably:
\begin{itemize}
	\item \textbf{BIP 32} (2012-02-11, Informational, Final): Hierarchical deterministic wallets are proposed that can be interchanged between clients, and allow generation of public keys without disclosing the private key.
	\item \textbf{BIP 42} (2014-04-01, Standards Track, Final): Introduction of static money supply for Bitcoin.
	\item \textbf{BIP 50} (2013-03-20, Informational, Final): A block that had a larger number of total transaction inputs than previously seen was mined and broadcasted. Bitcoin 0.8 nodes were able to handle this, but some pre-0.8 Bitcoin nodes rejected it, causing an unexpected fork of the blockchain. Bitcoin 0.8 nodes had to downgrade to resolve  the fork, during this time there was at least on large double spend, which was not intended to be malicius (i.e., was experimental).
	\item \textbf{BIP 141} (2015-12-21, Standards Track, Final): Defines a new structure called a "witness" that is committed to blocks separately from the transaction merkle tree. This structure contains data required to check transaction validity but not required to determine transaction effects. In particular, scripts and signatures are moved into this new structure. The witness is committed in a tree that is nested into the block's existing merkle root via the coinbase transaction for the purpose of making this BIP soft fork compatible.
	\item \textbf{BIP 341} (2020-01-19, Standards Track, Final): Proposes a new SegWit version 1 output type, with spending rules based on Taproot, Schnorr signatures, and Merkle branches.
\end{itemize}
We decided to construct a separate signal time series out of these major BIPs as we expect them to have the most significant impact on the cleaned wealth distribution data.\par
\subsubsection{All Economy-Related BIPs Expect the Major Economy-Related Ones}
We define this category as the set subtraction between the \emph{All Economy-Related BIPs} and \emph{Major Economy-Related BIPs} (Note that the latter is the subset of the former). The reason we included this set of BIPs into our analysis is because if the previous two categories produce deviating results, we would like to understand what exact difference do the \emph{Major Economy-Related BIPs} make, so this subset also acts as a control test just like the category \emph{All BIPs}.\par
\subsection{Granger-Casusality Test}
Having motivated our selection of BIPs for constructing the signal time series, we now move to the selection of the causality testing techniques available to us. The tool that is most commonly used for this purpose in the industry and acadmic research is the Granger-causality test \cite{GrangerCausalityTest}.\par
The method proposed by Granger \cite{GrangerCausalityTest} relies on breaking down the two time series, whose causal relationship is being tested, into their oscillator subparts using the Fourier transformation, which is a very theoretical approach and which can lead to long compute times and inaccuracies when used in computational data processing. For this reason, in statistical packages like \emph{statsmodels} \cite{StatsModels} there exist Granger-causality test functions that are optimized for computational analysis. Upon examination of different statistical packages the authors have found inconsistencies of implementation between different statistical packages and methodologies proposed as alternatives to the spectral analysis proposed by Granger in his original work \cite{GrangerCausalityTest}.\par
In online resources like Wikipedia \cite{GrangerCausalityTestWikipedia}, the process for testing for Granger-causality can be summarised as the follows given the independent variable \emph{X} and the dependent variable \emph{Y}:
\begin{enumerate}
	\item Create an autoregressive model of \emph{Y}, with the order of model being determined by the partial autocorrelation function, where the order is the highest order of \emph{Y} above some set critical level. Call this model a nested model.
	\item Extend the nested model to include autoregressive terms of \emph{X}, where the order of \emph{X} is selected by minimizing an information criterion (e.g., Akaike, Bayesian, etc.).
	\item Iteratively filter out the insignificant lagged terms of \emph{X} until only the terms that pass t-test at a given critical level remain. Call this model the parent model.
	\item Run the F-test between the nested and parent models to determine whether adding the independent variable \emph{X} explains away some of the variance, given the specified critical level for the F-test.
\end{enumerate}
Such procedure reasonably replaces the original procedure introduced by Granger. However, upon examining the Granger-causality test in the \emph{statsmodels} Python package, it became apparent that the test implemented there uses a variable called $max\_order$ which acts as the order for the lagged terms of \emph{X} and \emph{Y} simultaneously, therefore, creating nested and parent multivariate linear regression models with orders $O(max\_order)$ and $O(2max\_order)$ respectively. The Granger-causality test in the package also ignores the t-test filtering.\par
The problem with this implementation is that it is a lot less limiting than the one suggested by Wikipedia \cite{GrangerCausalityTestWikipedia}, and yet the package only cites the original paper, which takes a completely different approach. To ensure the validity of results, we decided to use both test variations calling the \emph{statsmodels} implementation a ``Simple'' test and the Wikipedia one a ``Full'' test, since it provides a more sophisticated filtering process. Both of these techniques are implemented in the Rust package DigiFi \cite{DigiFi}, which we used for the analysis.\par
\subsection{Endogenous Policies Analysis}
Now that we have selected the sets of BIPs and the tools that we will use in our analysis, we can perform the tests. From Section \ref{section:ImpactOfexogenousFactorsOnWealthDistribution}, we know that the cleaned wealth distribution data is in the monthly format as most of the exogenous factors we have used to clean data are quoted on a monthly basis. This means that when we convert the BIP sets we have selected into signal time series, we have to make it monthly as well (i.e., if there was a BIP relased in that month, the value of the time series for that month is $1$ and otherwise it is $0$). Appendix \ref{appendix:CleanedSupplyDistributionData} shows each cleaned supply distribution bucket in a blue line, and the BIPs as the dashed vertical red lines (Note that these plots include every BIP).\par
As there are ten cleaned wealth distribution buckets, four BIP signal time series and two Granger-causality test variants, we will have $80$ results, where the value $T$ (i.e., True) indicates the existence of Granger-causality between the respective BIP signal time series and the cleaned wealth distribution data, while $F$ (i.e., False) means there is not enough evidence to conclude that there is Granger-causality. For further convenience, we have added extra information about each of the tests' results - for tests that have shown a Granger-causality we have added a longest significant lag in the parent model in the brackets next to $T$ (note that for ``Simple'' test it will always be equal to the maximum lag that we set, as ``Simple'' test does no apply t-test filtering`), whereas for failed tests we have included the reason for its failure in the bracket next to $F$. Also note that we have used very standard parameters for tests: for the ``Full'' test we used $10\%$ critical interval for the partial autocorrelation function with its maximum lag set to $10$ (i.e., $10$ months in terms of our data), $10$ maximum lag for the independent variable, and $5\%$ critical level for both t-test and F-test; for the ``Simple'' test we have used $10$ maximum lag (as described above it applies to both time series) and a $5\%$ critical level for the F-test (``Simple'' test does not have t-test, so there is no critical level for it).\par
The results are presented in Appendix \ref{appendix:BIPsCausalityAnalysis}, where the cells with $T$ have been highlighted for convenience. The first thing we note is the consistency between the two Granger-causality test implementations, where we see a similarity between the patterns across the data. This is important as it means that different causality test implementations yield similar results.\par
The second observation is that, as expected, \emph{All BIPs} signal time series doesn't show significant causal relationship with any of the wealth distribution buckets as it potentially proves to be too uniform. Other signal time series, however, have all showed some causal relationship with at least one of the wealth distribution buckets.\par
Furthermore, we note that \emph{All Economy-Related BIPs} set seems to cause significant wealth redistribution amongst the wealthy addresses, yet  the \emph{Major Economy-Related BIPs} set has impacted the poorer and mid-wealth addresses. To further formulate the hypothesis that can help explain this pattern in the data, we must mention that there is a distinction between addresses and real people, and therefore, the behavioural factors that may impact one person may impact multiple addresses that are controlled by that individual and vice versa, when multiple people share an ownership of one address the behavioural factors may not impact that address, yet impact the people. Because of the lack of data, in this analysis we assume that the addresses correspond to people, and therefore, when we talk about the behavioural factors that may affect addresses we assume that these factors affect the people who are represented by these addresses.\par
The significant wealth redistribution among the wealthier addresses is an interesting pattern as it may suggest that the more wealthy addresses are more engaged in the Bitcoin ecosystem as they have higher stake in it, while the addresses with less wealth are less engaged in the process and they only respond to large economic events as these get propagated through many different communication and marketing channels. It could also signify that the more wealthy addresses better anticipate the major changes (or even cause them as the wealthy addresses are the de facto governors of the system) and they gradually move wealth, therefore, pricing-in the major changes to the Bitcoin economy. To add to this from a different perspective, it may also imply that if the majority of poorer addresses are owned by a small group of individuals, it may point to the difficulty of redistributing wealth by these individuals when the small changes to the economy are made, whereas when the big policy changes are introduced, they may be forced to react. This, however, does not change the pattern in the data: the wealthier addresses seem to respond to the ``minor'' economic policy changes, and the poorer addresses seem to be more reactive to the ``major'' policy changes.\par
In general, we hypothesize that this discrepancy is caused by the information asymmetry where the wealthy addresses move wealth quickly in response to the release of ``minor'' economic policies as they constantly monitor the space and have the resources to do so, yet they price in the ``major'' economic policy changes as they either directly cause these changes or are aware of them. On the contrary, the owners of the poorer addresses seem to either pay less attention or have poor information about the ``minor'' economic policy changes so they don't respond to it, while the ``major'' policy changes cause poorer addresses to make reactive investment decisions as they don't anticipate the changes. We can visualize this hypothesis by the table \ref{table:ImpactOfEndogenousPoliciesOnWealthDistribtuion:HypothesisVisualization}, where the lack of the response to the policy changes is indicated by $F$ and statistically significant response (i.e., significant log-change in wealth distribution) to policy changes is indicated by $T$.\par
\begin{table}[h!]
\centering
\begin{tabular}{ |c|c|c| } 
	\hline
	& \textbf{Wealthier Addresses} & \textbf{Poorer Addresses} \\
	\hline
	\begin{tabular}{@{}c@{}}\textbf{Major Economic} \\ \textbf{Policy Changes} \end{tabular} & \begin{tabular}{@{}c@{}}False \\ (Priced In) \end{tabular} & True \\
	\hline
	\begin{tabular}{@{}c@{}}\textbf{Minor Economic} \\ \textbf{Policy Changes} \end{tabular} & True & \begin{tabular}{@{}c@{}}False \\ (Little Resources/Small Stake) \end{tabular} \\
	\hline
\end{tabular}
\caption{Possible explanation of patterns present in causation analysis between BIPs and wealth redistribution}
\label{table:ImpactOfEndogenousPoliciesOnWealthDistribtuion:HypothesisVisualization}
\end{table}
Next, we see that the control tests with \emph{All Economy-Related BIPs Except Major Economy-Related Ones} signal time series shows a pretty similar picture to the tests run with the \emph{All Economy-Related BIPs} as expected with only difference being the ``Simple'' test with \emph{From 10 to 100} bucket. The possible reason for this is that the test value was very close to the critical level and it barely managed to pass, but we do not see a trend that would suggest that removing \emph{Major Economy-Related BIPs} from \emph{All Economy-Related BIPs} should produce a completely different set of results.\par
Finally, we must discuss the sensitivity of our results to the parameters that we have used. In the table in Appendix \ref{appendix:BIPsCausalityAnalysis}, we see that for the ``Full'' test there is a cell that implies that the lag of $10$ months is still significant, so we wondered what happens if we shrink or extend the test window. We decided to re-run the same Granger-causality tests, but this time with $6$ and $12$ as the maximum lag for the partial autocorrelation function and independent variable (i.e., $6$ and $12$ months), the results of these tests are presented in Appendices \ref{appendix:BIPsCausalityAnalysis6Months} and \ref{appendix:BIPsCausalityAnalysis12Months} respectively. Despite the slight difference in the results, the overall pattern seems to remain the same, which makes us believe that this  pattern is relatively stable and is not significantly impacted when changing the test window. The other interesting result from this sensitivity analysis, is that in the $12$ months table in Appendix \ref{appendix:BIPsCausalityAnalysis12Months}, the ``Full'' test shows that the longest significant lag is $6$ months, which lines up with the result from the $6$ months table (note that despite not being presented in the table in Appendix \ref{appendix:BIPsCausalityAnalysis}, the cell that shows $10$ months as the longest significant lag for the ``Full'' test also has $6$ months lag as being significant according to the t-test, so all three tables line up in that conclusion). Therefore, we hypothesise that $6$ months is the longest significant period for BIPs to make an impact on wealth distribution as per the results from the ``Full'' test, while the $10$ months cell we see as a statistical anomaly.
\subsection{Summary}
\label{subsection:ImpactOfEndogenousPoliciesOnWealthDistribtuion:Summary}
These results are interesting because they highlight potential patterns in wealth redistribution, but it also signifies that our approach of selecting policies, albeit somewhat subjectively, is not a coincidence and that we potentially can formalize it into a taxonomy. We also see that BIPs do have an impact on wealth distribution, and that different sets of policies have impacts on different wealth distribution buckets in the Bitcoin economy. This section, in conjunction with the previous one, proposes a technique for analysis of policies and their impact on token economies, and can potentially be generalized to standard economies given the appropriate data.\par
This methodology can be further improved by adding a more in-depth analysis of the linear regressions that were constructed as part of the ``Full'' Granger-causality test such as the review of the p-values produced by the iterative t-tests for every coefficient and the p-value of the F-test. This would explain the deviation between the ``Full'' and ``Simple'' tests, and provide feedback on whether the selected confidence levels were too restrictive or permissive. The other improvement that could be added to this analysis technique is a procedure for determining the maximum order of the partial autocorrelation function and maximum lag of the independent variable in the ``Full'' test, and the maximum lag in the ``Simple'' test. For the purpose of the paper, we assume that BIPs do not have an impact delay of over $10$ months and from the sensitivity analysis we see that the most significant impact delay is $6$ months, but it would be useful to define a procedure for determining that impact delay window. However, we leave both of these improvement proposals for future work.

\section{Economic Policy Taxonomy of Token Economies}
\label{section:PolicyTaxonomyOfTokenEconomies}
In this section we aim to formalize the observation made in Subsection \ref{subsubsection:AllEconomyRelatedBIPs}, where we have clustered BIPs into categories justifying it by stating that they seem to be ``economy-related''. To create a taxonomy of economic policy in token economies, we will use the frameowrk proposed in \emph{Economic Policy Taxonomy }\cite{EconomicPolicyTaxonomy} as the backbone, and will map BIPs to parts of that framework, and vice versa, we will use that framework to suggest possible policies that could be implemented in token economies. However, upon previous examining of BIPs we know that some of these policies, despite us considering them economic policies, cannot be mapped directly to the policies in standard economies. For this reason we will split this section into two parts: the first part is the taxonomy of economic policies in token economies that can be mapped to the policies in standard economies; and the second part is the taxonomy of economic policies that are specific to the token economies.\par
Before we begin, notice that here we propose a starting point for a taxonomy that should be extended and modified with future research, rather than accepted as the complete and exhaustive list. In this section we adopt the definitions of policy and atomic policy as per \emph{Economic Policy Taxonomy} \cite{EconomicPolicyTaxonomy}:
\begin{definition}
	\textbf{Policy} is a set of ideas or a plan of what to do in particular situations that has been agreed to officially by a group of people, a business organization, a government or a political party.
\end{definition}
\begin{definition}
	\textbf{Atomic Policy} is an individual action or idea that applies a change to a specific body or phenomena.
\end{definition}
We specifically aim to identify atomic policies as these are the individual actions that cannot be broken down into smaller set of actions. These are very useful to understand as a global economic policy is a set of atomic actions implemented at once, which are aimed at specific parts of the economy. Understanding possible atomic policies means we are able to design more precise economy models using the DeTEcT framework.\par
\subsection{Standard Economic Policy Taxonomy in Token Economies}
The caveat of recreating a policy taxonomy framework for token economies is that we need a government and central bank to map the atomic policies from standard economies, but since token economies are considered decentralised there is no one agent in the economy that is responsible for setting policies and redistributing wealth. To solve this issue we can use \emph{Control Mechanism} proposed in DeTEcT \cite{DeTEcT} as an agent in the economy that receives and spends wealth, which we can think of as the protocol itself. While it is true that the protocol itself is goverened by other agents in the economy, and therefore, there is a distinction in governance and wealth redistribution flows, the taxonomy is only concerned with the cash flows and not the governance infrastructure, which without loss of generality can be abstracted away.\par
Since \emph{Control Mechanism} is considered to be a unique agent, yet in standard economy we would have \emph{Government} and \emph{Central Bank} as two different entities, we propose preserving the distinction between fiscal and monetarty policies even in token economies as it will provide a better mapping with the standard economy policies and will allow us in the future to separate the roles of \emph{Control Mechanism} into a ``government-like'' and ``central bank-like'' agents inside a token economy, which itself can be considered an atomic policy.\par
We start by introducing an income statement for \emph{Control Mechanism} where we separate it into two parts as described above: fiscal-like and monetary-like parts.
\begin{itemize}
	\item \textbf{Fiscal-Like Income Statement}
	\begin{itemize}		
		\item \textbf{Revenue}
		\begin{itemize}
			\item \textbf{Transaction Fees}: Fees paid directly to the protocol (Not to confuse with the transaction fees that are paid to the miners for including a transaction in a block).
		\end{itemize}
		\item \textbf{Expenses}
		\begin{itemize}
			\item \textbf{Cost of Sales}: Tokens issued to the agents that maintain economy operation (e.g., miners).
			\item \textbf{Incentives}: Incentives issues by \emph{Control Mechanism} to economy participants for satisfying a predetermined condition.
		\end{itemize}
	\end{itemize}
	\item \textbf{Monetary-Like Income Statement}
	\begin{itemize}
		\item \textbf{Revenue}
		\begin{itemize}
			\item \textbf{Sale of Foreign Currency}: Reaquisition of tokens for other type of currency.
		\end{itemize}
		\item \textbf{Expenses}
		\begin{itemize}
			\item \textbf{Purchase of Foreign Currency}: Sale of tokens for other type of currency.
		\end{itemize}
	\end{itemize}
\end{itemize}
The statements defined here are a lot more compact than the ones defined in \emph{Economic Policy Taxonomy} \cite{EconomicPolicyTaxonomy} as there are fewer moving parts in token economies than there are in standard economies. However, as there are limited  number of BIPs, we can more clearly outline atomic policies that \emph{Control Mechanism} can use to regulate an economy and provide examples of policy implementations. Below is the mapping between the token economy fiscal-like policies and a standard economy fiscal policies:
\begin{itemize}
	\item \textbf{Revenue Policy}
	\begin{itemize}
		\item \textbf{Transaction Fee Size}: Amount of tokens payable for transaction completion that will be ``burned'' (i.e., Removed from current circulation).
	\end{itemize}
	\item \textbf{Expenditure Policy}
	\begin{itemize}
		\item \textbf{Mining Fee Size}: Size of rewards issued by the protocol to the agents that produce and validate transaction blocks.
		\item \textbf{Intentive Size}: Size of reward issued by the protocol to the agents that satisfy specific criteria.
		\item \textbf{Incentive Criteria}: Criteria that must be satisfied by an economy participant in order to receive a token reward.
	\end{itemize}
	\item \textbf{Auxiliary Policy}
	\begin{itemize}
		\item \textbf{Auctions}: Allowing for economy participants to negotiate the terms of a transaction among themselves (e.g., \emph{BIPs 78, 199}).
	\end{itemize}
\end{itemize}
Since most policies implemented by \emph{Control Mechanism} are directed at changes in supply or regulation of transactions and trade of financial instruments inside a token economy, there are far more monetary-like atomic policies that we have managed to identify in using BIPs:
\begin{itemize}
	\item \textbf{Debt and Credit Policy}: Policy surrounding creation or issuance of debt by the \emph{Control Mechanism}.
	\begin{itemize}
		\item \textbf{Fixed Supply}: Maximum bound for the supply set in the economy (e.g., \emph{BIP 42}).
		\item \textbf{Dynamic Supply}: Infinite maximum bound for the supply set in the economy.
		\item \textbf{Supply Increment}: Additional issuance of tokens to the maximum supply.
		\item \textbf{Supply Decrement}: Removal of tokens from the maximum supply.
	\end{itemize}
	\item \textbf{Financial Markets Policy}: Regulation of transactions and financial instruments issued or build on top of the token economy.
	\begin{itemize}
		\item \textbf{Complex Transaction}: Transactions with conditions attached that are defined by the transacting parties that describe input, output or intermediary constraints that have to be satisfied for the transaction to be successful (e.g., \emph{BIPs 13, 16}).
		\begin{itemize}
			\item \textbf{Number of Signatures}: Number of signatures required by the transaction to complete (e.g., \emph{BIP 11}).
			\item \textbf{Relative Lock-Time}: Timestamp that triggers the completion of transaction (e.g., \emph{BIP 68}).
			\item \textbf{Future Agreement}: The output of a completed transaction is unspendable until some timestamp (e.g., \emph{BIP 65}).
			\item \textbf{Debt Agreement}: Transaction contract that creates a debt agreement between the transacting parties (e.g., \emph{BIP 197}).
			\item \textbf{Transaction Order}: Order in which consequtive transactions must be executed.
			\item \textbf{Proof of Identity}: Request to obtain a proof of identity before the completion of the transaction (e.g., \emph{BIP 75}).
			\item \textbf{Proof of Reserves}: Request to obtain proof of reserves before the completion of the transaction (e.g., \emph{BIP 127}).
			\item \textbf{Destination}: Restrictions on the future destinations of the transaction (e.g., \emph{BIPs 112, 118}).
			\item \textbf{Recovery Path}: Specification of transaction recovery path that can be used to recover wealth to the specified address before the transaction is complete (e.g., \emph{BIP 345}).
			\item \textbf{System State}: State of economy or system is used to constrain the transaction (e.g., \emph{BIP 115}).
			\item \textbf{Mutable Transactions}: Ability for a transaction participant to mutate the state of the transaction after its completion (e.g., \emph{BIPs 125, 141, 370, 371, 373}).
			\item \textbf{Offline Transactions} Support for offline and/or delayed transactions (e.g., \emph{BIP 174}).
			\item \textbf{Ancestor Package Propagation}: Request to relay an unconfirmed ancestor package of the given transaction for in-depth filtering of future transaction tree (e.g., \emph{BIP 331}).
		\end{itemize}
		\item \textbf{Transaction Validation}: Validation of transactions against rules defined by the protocol (e.g., \emph{BIPs 53, 54, 66, 146, 173}).
		\begin{itemize}
			\item \textbf{Double Spend Validation}: Validation for the transaction not to spend the wealth that has been spent already or mimic another transaction by id or other metadata (e.g., \emph{BIP 30}).
			\item \textbf{Transaction Size}: Maximum cap on the size of the transaction to be included in the block and, by extension, the block itself (e.g., \emph{BIPs 100, 101, 102, 103, 104, 105, 106, 107, 109}).
			\item \textbf{Metadata}: Standardisation and requirements for the transaction metadata that is passed into the block or between peers (e.g., \emph{BIP 152}).
			\item \textbf{Withdrawal Validation}: Requirements for the withdrawal of wealth from L2 and higher layer solutions to the L1 (e.g., \emph{BIP 300}).
		\end{itemize}
	\end{itemize}
\end{itemize}
This concludes the initial mapping between the atomic policies we identified in token economies to the atomic policies defined in standard economies. Note that the general structure of the taxonomy has been well-preserved, which hints at some overlap between policy making in standard economies versus token economies.\par
\subsection{Economic Policy Taxonomy Specific to Token Economies}
In this section we cover the economic policies that are specific to the token economies, and therefore, didn't make it into the taxonomy defined in the previous section. Note that their exclusion from the former taxonomy part does not say anything about their potential impact on the wealth distribution.\par
\begin{itemize}
	\item \textbf{Liquidity Policy}: Policies directed at the change in liquidity or transaction throughput in the economy.
	\begin{itemize}
		\item \textbf{Hard Fork}: Split of the blockchain protocol by nodes based on the applied set of policies.
		\item \textbf{Simultaneous Soft Forks}: Ability of a blockchain to run multiple soft forks simultaneously without loosing liquidity due to different software versions (e.g., \emph{BIP 9}).
		\item \textbf{Fork Activation Requirements}: Requirements laid out by the economy participants that trigger the activation of a fork (e.g., \emph{BIPs 91, 148, 343}).
	\end{itemize}
	\item \textbf{Wealth Storage Policy}: Policy that describes how and where wealth will be stored (e.g., wallets)
	\begin{itemize}
		\item \textbf{Hierarchical Deterministic Wallets}: Wallets that generate addresses in a hierachical structure of addresses that do not disclose a private key and can be shared between economy participants for the purpose of transacting and storing wealth (e.g., \emph{BIPs 32, 44, 175}).
		\item \textbf{MultiSig Wallets}: Wallets that require multiple private keys to sign off a transaction (e.g., \emph{BIP 129}).
		\item \textbf{Aggregate Key}: Allows generation of aggregate key by multiple signers (e.g., \emph{BIP 327}).
		\item \textbf{Wallet Constraints}: Constraints imposed on wallets (e.g., \emph{BIP 88}).
	\end{itemize}
	\item \textbf{Social Policy}: Policies directed at simplifying, popularizing or enhancing the adoption of token economy by general public (e.g., \emph{BIPs 61, 380}).
	\begin{itemize}
		\item \textbf{Standardized Policy Design Mechanism}: Mechanism that enables economy participants to create improvement proposals, vote on them and implement them (e.g., \emph{BIP 1}).
		\item \textbf{Payment Schemes}: Mechanisms for facilitation of transactions between vendors and customers (e.g., \emph{BIPs 21, 47, 70, 72, 79, 111, 322, 324, 330, 341, 351, 352}).
		\item \textbf{Mnemonic Schemes}: Design policy to allow the use of easy-to-remember variables such as primary or public keys by economy participants (e.g., \emph{BIP 39}).
	\end{itemize}
	\item \textbf{Auxiliary Policies}: Policies that are unique and have limited overlap with other policies.
	\begin{itemize}
		\item \textbf{Protocol Choice}: Choice of protocol and its rules for processing and validating transactions, as well as allowing the development of higher layer solutions.
		\item \textbf{Monetary Authority}: Unique protocol for regulating wealth redistribution and controlling supply vs separate governance and fiscal institutions.
	\end{itemize}
\end{itemize}
\subsection{Analysis of Policy Taxonomy Compartments}
Having proposed a taxonomy, we have allocated BIPs from the \emph{All Economy-Related BIPs} set to the respective atomic policies or parts of the taxonomy that need to be broken down into atomic policies in the future work (e.g., \emph{Auctions}). Our proposed taxonomy contains three major policy categories: \emph{Fiscal-Like Policies}, \emph{Monetary-Like Policies} and \emph{Purely Tokenomic Policies} (i.e., Policies specific to the token economies). Since we have BIPs reprsenting each of these categories in our taxonomy, we decided to see if the Granger-causality analysis described in Section \ref{section:ImpactOfEndogenousPoliciesOnWealthDistribtuion} can reveal any patternss in this categorisation, so we have run the Granger-causality tests again but with the following three sets of BIPs:
\begin{itemize}
	\item \textbf{Fiscal-Like BIPs}: 78, 199
	\item \textbf{Monetary-Like BIPs}: 11, 13, 16, 30, 42, 53, 54, 65, 66, 68, 75, 100, 101, 102, 103, 104, 105, 106, 107, 109, 112, 115, 118, 125, 127, 141, 146, 152, 173, 174, 197, 300, 331, 345, 370, 371, 373
	\item \textbf{Purely Tokenomic BIPs}: 1, 9, 21, 32, 39, 44, 47, 61, 70, 72, 79, 88, 91, 111, 129, 148, 175, 322, 324, 327, 330, 341, 343, 351, 352, 380
\end{itemize}
The parameters for the tests were as in the section before: for the ``Full'' test - $10\%$ critical interval for the partial autocorrelation function with maximum lag of $10$, $10$ maximum lag for the independent variable, and $5\%$ critical level for the t-test and F-test; for the ``Simple'' test - maximum lag of $10$, and $5\%$ critical level for the F-test. The results of the analysis are presented in Appendix \ref{appendix:TaxonomisedBIPsCausalityAnalysis}, where $T$ indicates that the set of BIPs Granger-causes the change in the Bitcoin wealth distribution bucket.\par
From the table, we see that the results are very sporadic and don't showcase any visible patterns. The ``Simple'' and ``Full'' tests don't seem to be consistent, and there is also no clear division in what policies impact what wealth distribution buckets. One possible pattern we may trace is that \emph{Purely Tokenomics BIPs} set impacts a multitude of wealth distribution buckets.\par
A possible explanation of the discrepancy between this result and the causality analysis in Appendix \ref{appendix:BIPsCausalityAnalysis} is that different wealth buckets may pay attention to BIPs that introduce changes to the Bitcoin economy, but they do not care whether these are fiscal, monetary, or any other. We speculate that the responses we see in the wealth distribution buckets are caused by specific atomic policies rather than an implementation of a fiscal or monetary policy specifically, which is what we would expect to see.
\subsection{Summary}
In this section we devised a starting point for economic policy taxonomy of token economies. We have used the structure introduced for standard economies \cite{EconomicPolicyTaxonomy} to understand what atomic policies in token economies have their alternatives in standard economies and what policies are exclusive to token economies. This destinction can help us understand the structural difference between the token and standard economies. We analysed whether the attribution of a policy to fiscal or monetary policy makes a difference for the economy participants, and from our results we conclude that such labels don't seem to have consistent patterns for Bitcoin wealth redistribution. This taxonomy can be further enhanced by going though improvement proposals for token economies sush as Ethereum (i.e., EIPs), but we leave this for future work.

\section{Future Work}
Having conducted our research, we would like to point out a few areas for future research and development of the methodology proposed here. First, we would like to mention that there are many additional exogenous factors that we could consider filtering out in the future, such as factors outside the US and UK jurisdictions (e.g., ECB policies, Chinese central bank policies, etc.), fundamental shocks (e.g., world news, recessions, etc.) and many more. For this purpose, it would be useful to develop a framework that can be used to tell what factors need to be filtered out so that the economic time series in question can be considered ``cleaned''.\par
Second, the methodology for analysing the impacts of policies that we have suggested in this paper can be further improved as per subsection \ref{subsection:ImpactOfEndogenousPoliciesOnWealthDistribtuion:Summary}. Namely, more rigorous analysis of ``Full'' Granger-causality test can be derived, where the comparison of the linear regression models of the ``Full'' and ``Simple'' tests can be made to determine what lagged features have been filtered out and how sensitive the results of the Granger-causality test are to the changes in the confidence levels of t-tests and F-tests. In addition, a method for determining maximum lag parameters for both tests would be a good addition to the analysis methodology.\par
The results from Section \ref{section:ImpactOfEndogenousPoliciesOnWealthDistribtuion} can be examined further from perspective of token economies, but also from perspective of standard economies and finance. From perspective of token economies, we can try to prove the hypothesis that the wealthier addresses price in the changes in policy by examining the movement of wealth of individual wealthy addresses, while also examining when the poorer addresses tend to execute transactions in comparison to \emph{Major Economy-Related BIPs} signal time series. The time frame of transaction execution added to this research can potentially reveal further patterns in response patterns of different economy participants to the changes in economic policies.\par
However, these results may also hint at some patterns in standard economies and finance, which can be examined by applying the methodology descibed in this paper. For instance, we can check if ``major'' economic policy changes proposed or made by a government or central bank are indeed priced in by other economic participants, and whether information asymmetry does cause the discrepancy in response and actions of different Gini coefficient buckets.\par
Lastly, the economic policy taxonomy for token economies can be further improved by examining the policies implemented or proposed in other token economies such as Ethereum or Solana. These will expand our understanding of possible actions that the policy makers have, and will create a better reference for future research.

\section{Conclusion}
The aim of this paper was to analyse the impact of endogenous policies, represented by BIPs, on the wealth distribution in the Bitcoin token economy, check whether these policies have any impact on the distribution, and understand what atomic polices can be implemented by the regulators in an economy.\par
We have analysed what exogenous factors impact the Bitcoin wealth distribution, and we conclude that only \emph{Federal Funds Rate} and \emph{M2 (US)} have significant impact on \emph{From 0 to 0.001} and \emph{From 10 to 100} buckets of wealth distribution respectively. Using these results, we have cleaned the wealth distribution data from the noise produce by the external factors and proceeded to examine the causality between the endogenous policies (i.e., BIPs) and the cleaned wealth distribution. From this analysis we conclude that BIPs indeed Granger-cause wealth redistribution, and the poorer addresses in the Bitcoin token economy seem to be impacted by \emph{Major Economy-Related BIPs}, while the richer addresses are impacted by \emph{All Economy-Related BIPs}. This result may hint at the difference in attention that different economy participants pay to the policies being implemented due to the difference in ``stake'' (i.e., wealth) they have in the economy, or could be attributed to richer addresses being better at predicting major policy changes. According to the sensitivity analysis of the Granger-causality test, we also claim that the impact delay window for BIPs is $6$ months, so if new policies are being introduced they should take at most $6$ months to make a statistically significant impact.\par
Guided by the structure intoduced in the \emph{Economic Policy Taxonomy}, we constructed a taxonomy of policies in token economies. We motivated the choices of atomic policies with examples from BIPs, which explains their previous inclusion in the causality analysis. We note that the structure of the taxonomy we constructed for token economies is alike to the structure of policy taxonomy for standard economies, but there are some atomic policies that we could not map to the standard economies, which potentially demonstrates the structural difference in regulation of standard and token economies.\par
In conclusion, this paper has studied the impacts of exogenous and endogenous factors on the wealth distribution in Bitcoin ecosystem, introduced a starting point for categorization of economic policies in token economies, and proposed a procedure for the analysis of impacts of policies on wealth distribution.

\footnotesize % Font size for generic text and tables in appendix

\newpage
\appendix
\appendixpage
\section{Cleaned Supply Distribution Data}
\label{appendix:CleanedSupplyDistributionData}
\begin{itemize}
\setlength\itemsep{-0.5em}
	\item [\textbf{Hint}]
	\item Red dotted lines represent BIPs
	\item Year labels on the charts mark the start of the year
\end{itemize}
\begin{figure}[ht!]
	\centering
	\includegraphics[width=1\linewidth]{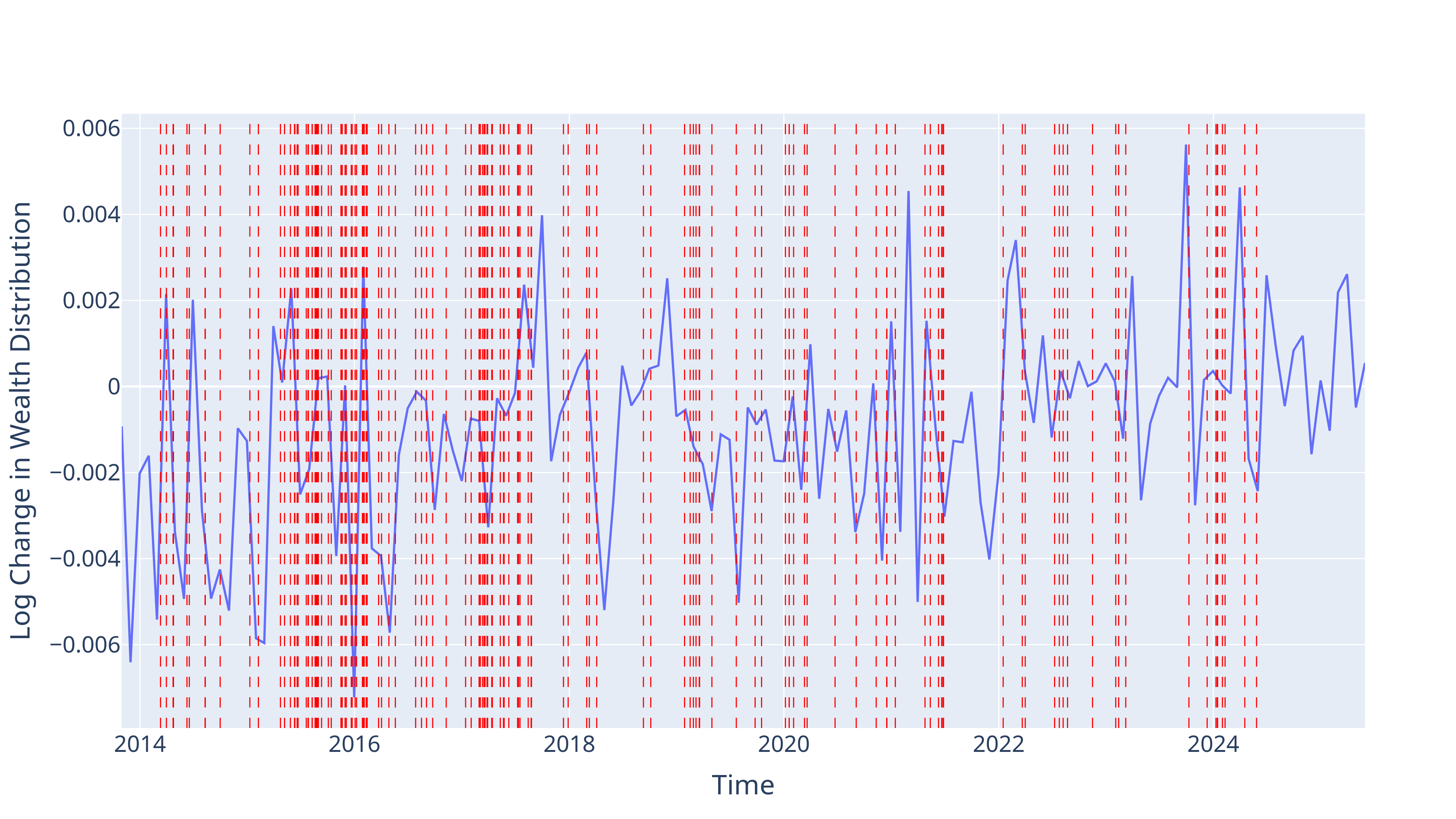}
	\caption{Log change in wealth distribution of bucket \emph{From 0 to 0.001}}
	\includegraphics[width=1\linewidth]{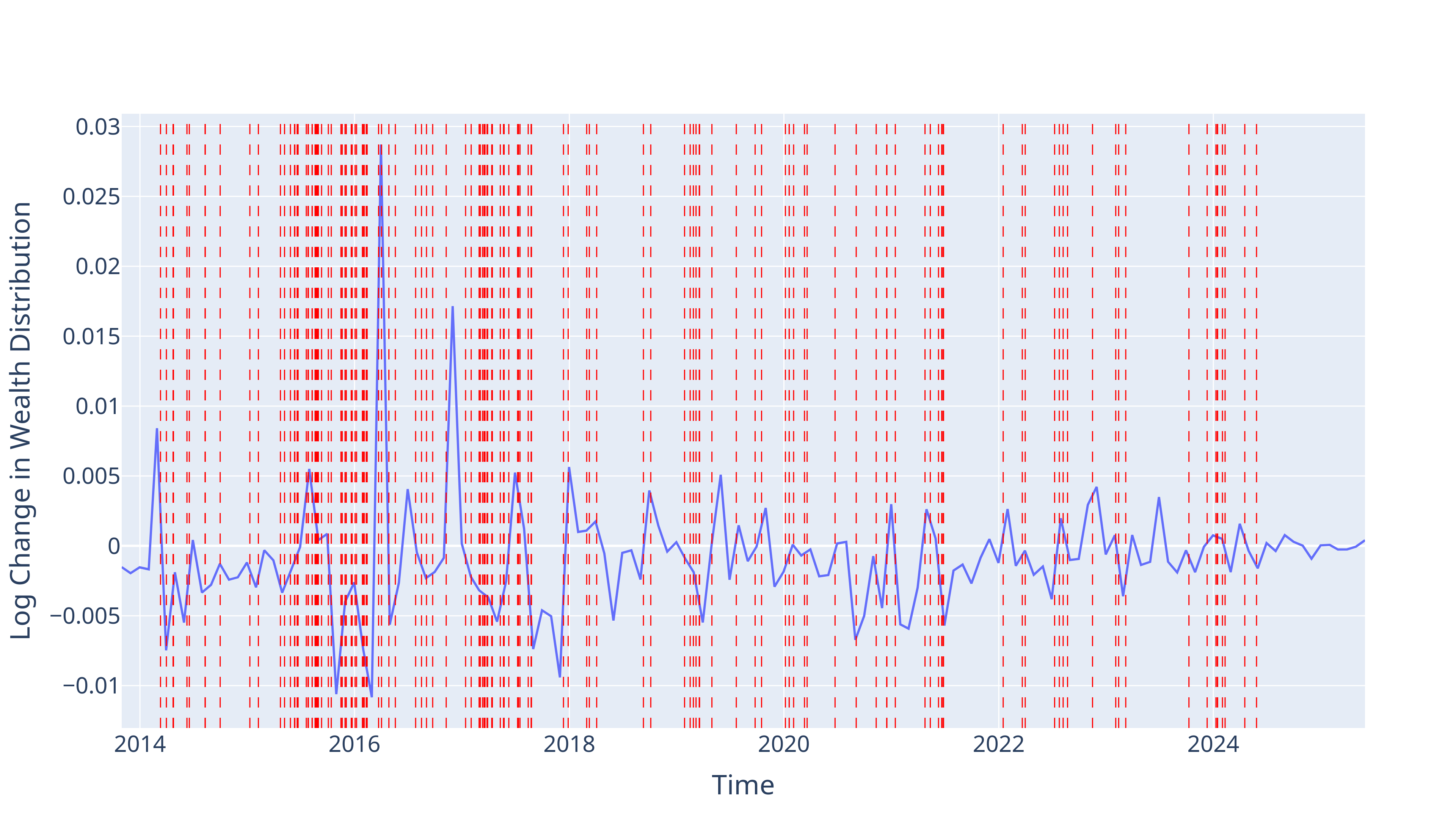}
	\caption{Log change in wealth distribution of bucket \emph{From 0.001 to 0.01}}
\end{figure}
\begin{figure}[ht!]
	\centering
	\includegraphics[width=1\linewidth]{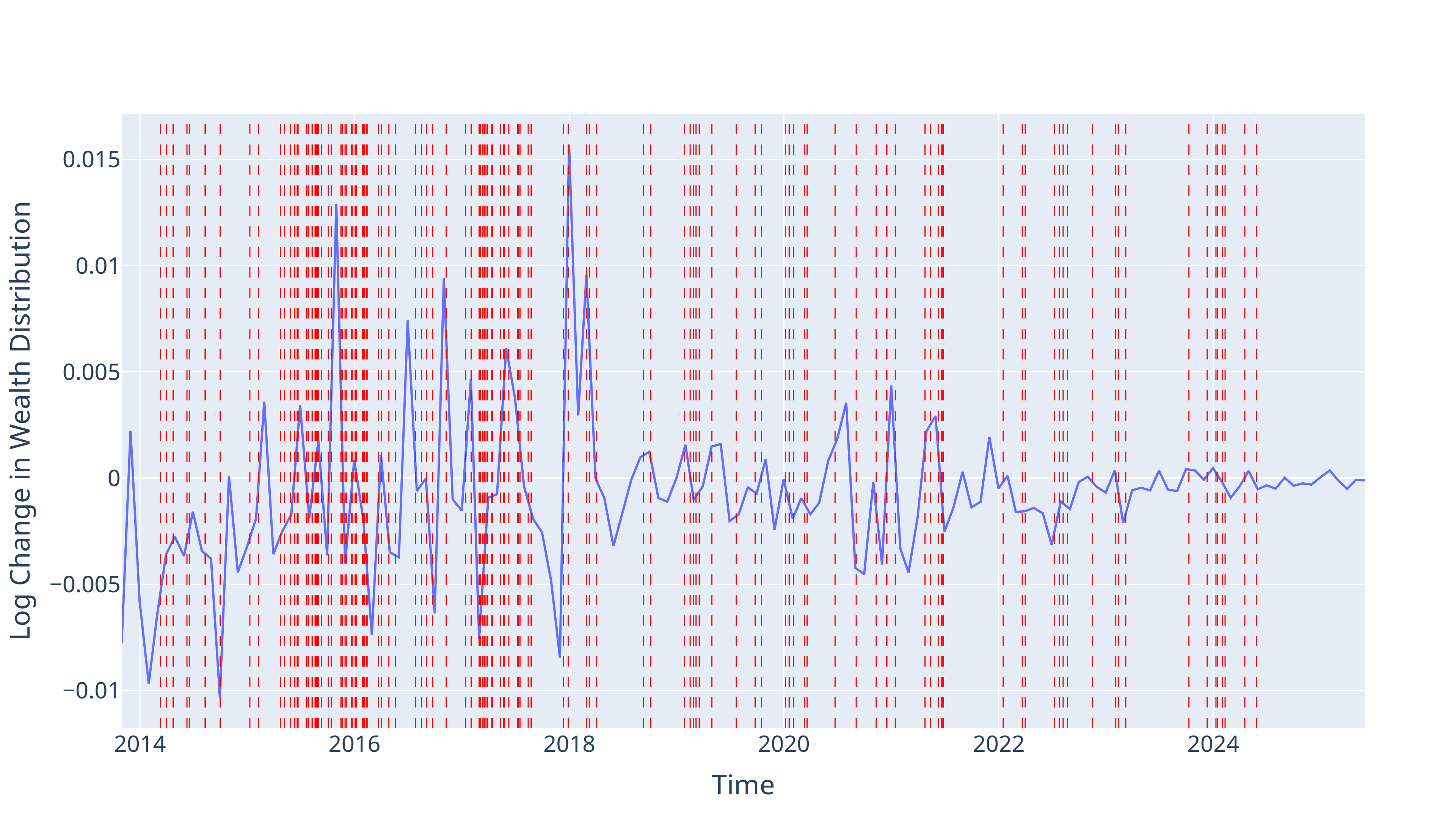}
	\caption{Log change in wealth distribution of bucket \emph{From 0.01 to 0.1}}
	\includegraphics[width=1\linewidth]{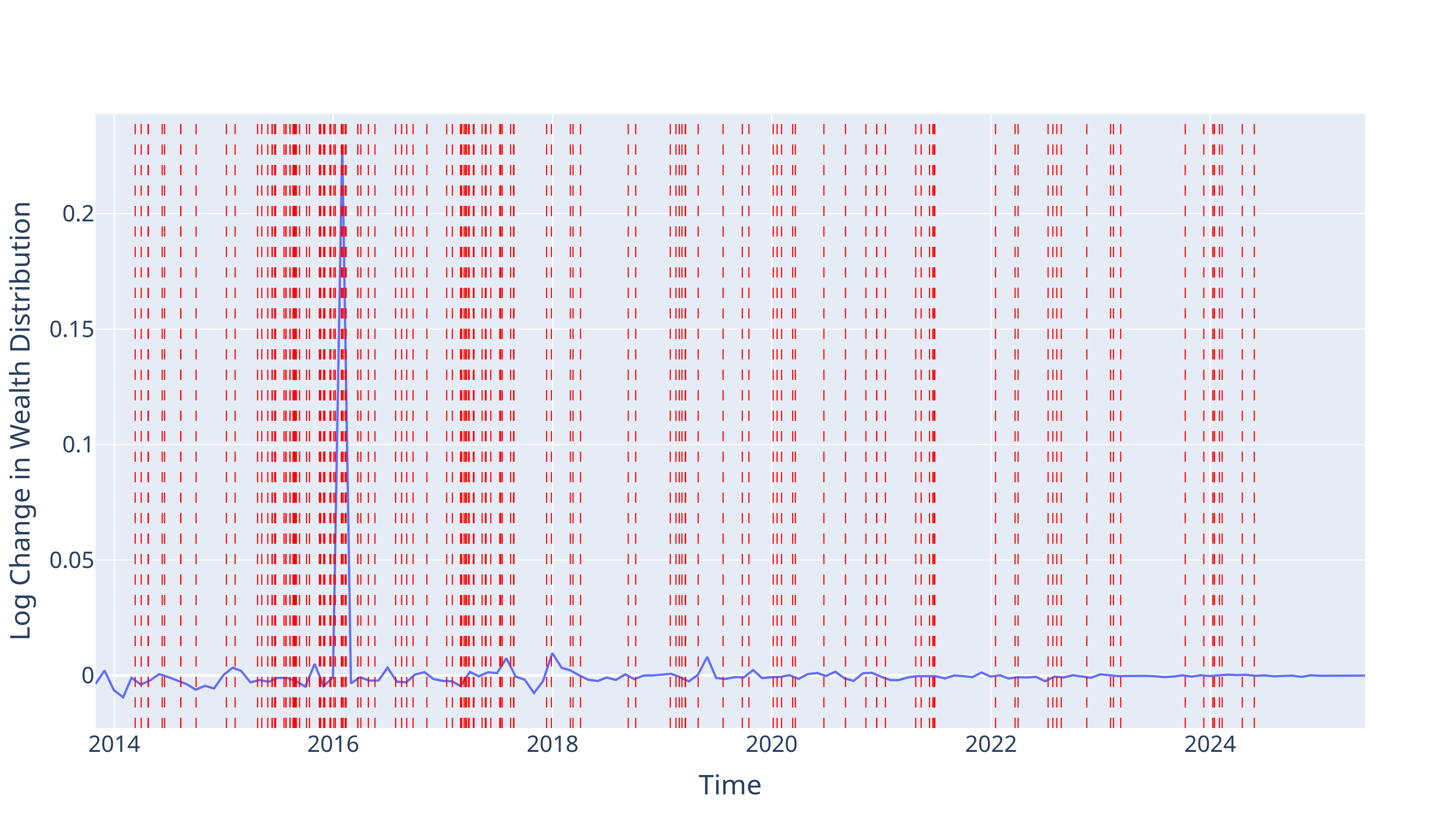}
	\caption{Log change in wealth distribution of bucket \emph{From 0.1 to 1}}
\end{figure}
\begin{figure}[ht!]
	\centering
	\includegraphics[width=1\linewidth]{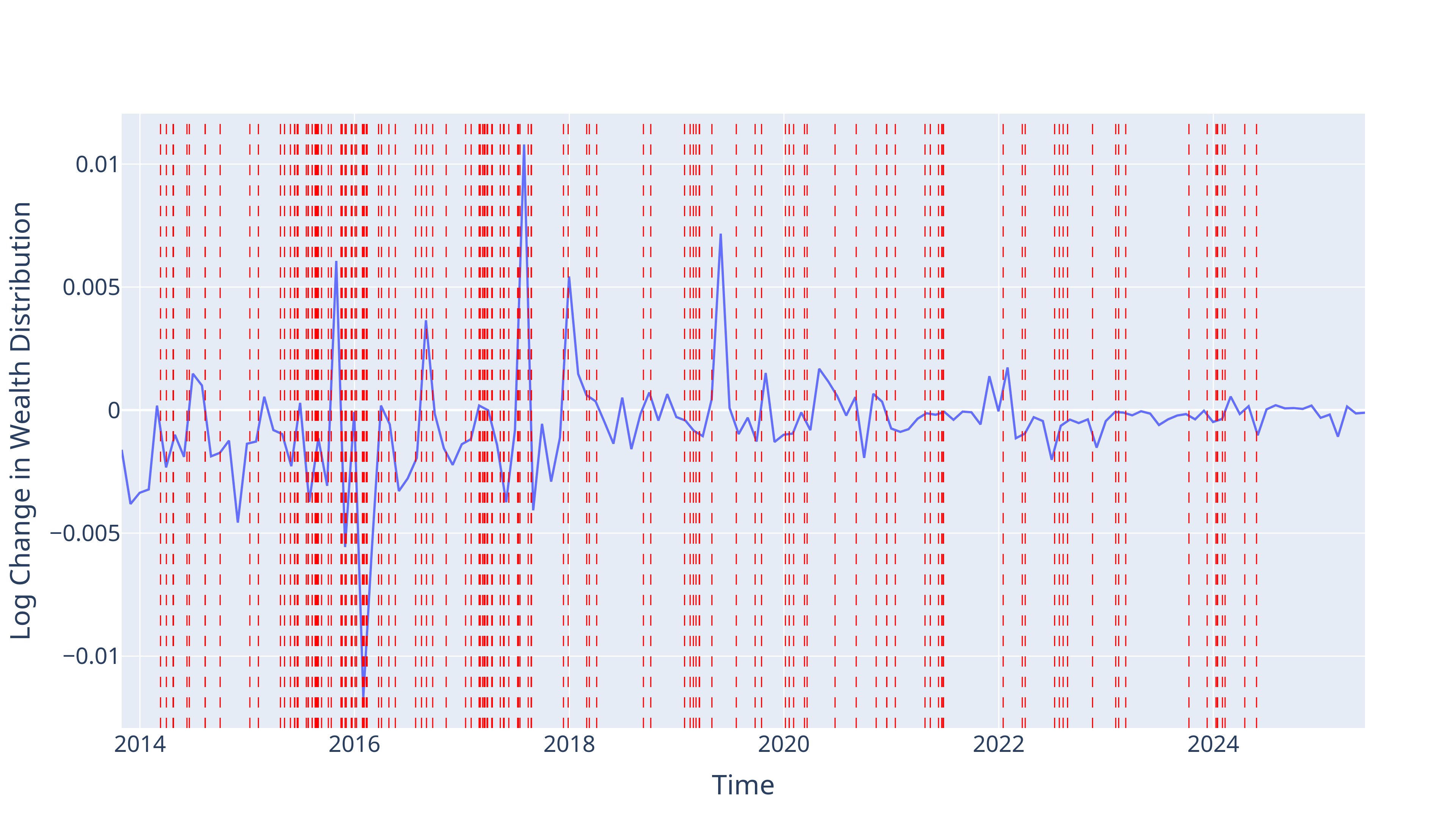}
	\caption{Log change in wealth distribution of bucket \emph{From 1 to 10}}
	\includegraphics[width=1\linewidth]{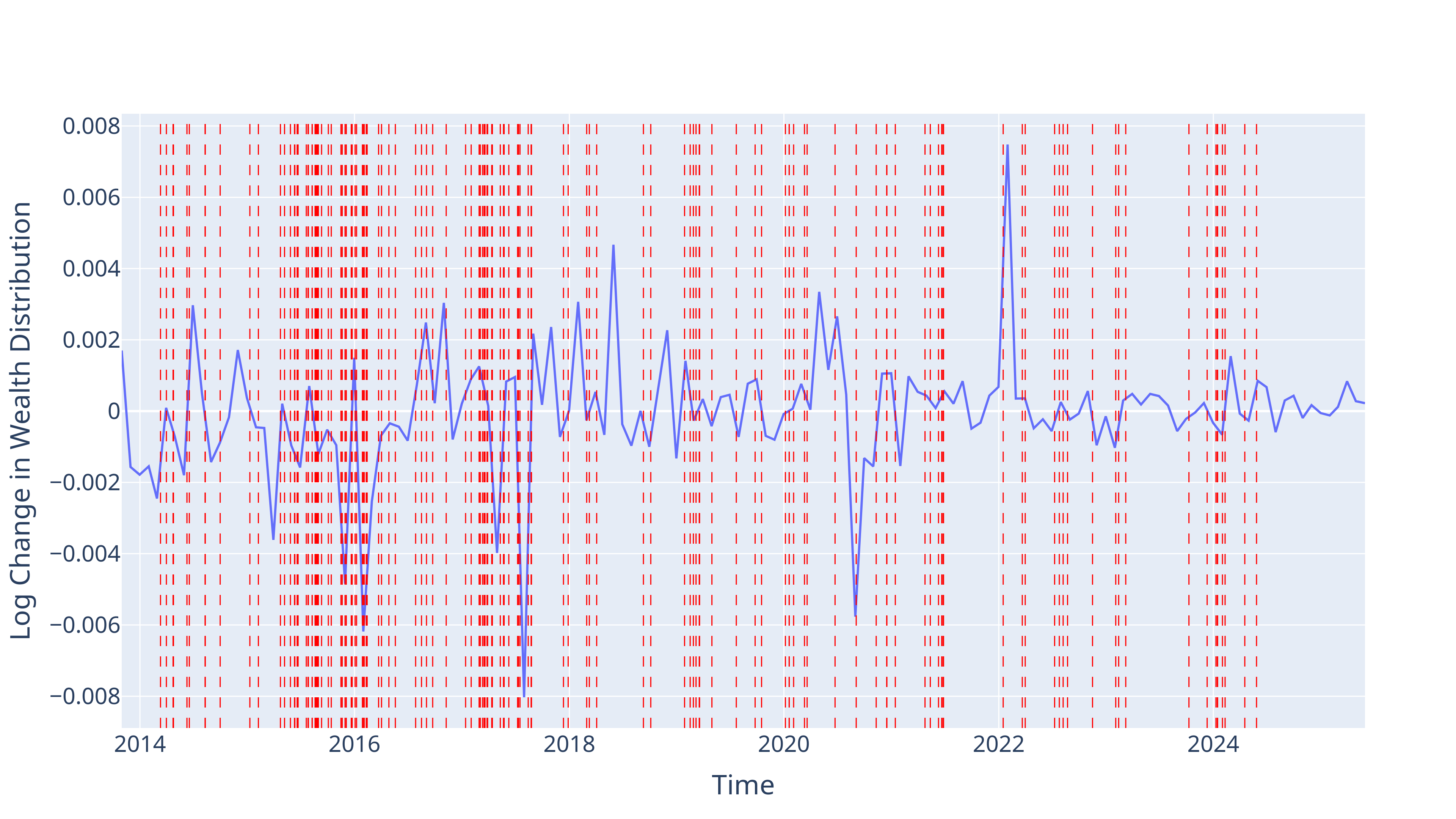}
	\caption{Log change in wealth distribution of bucket \emph{From 10 to 100}}
\end{figure}
\begin{figure}[ht!]
	\centering
	\includegraphics[width=1\linewidth]{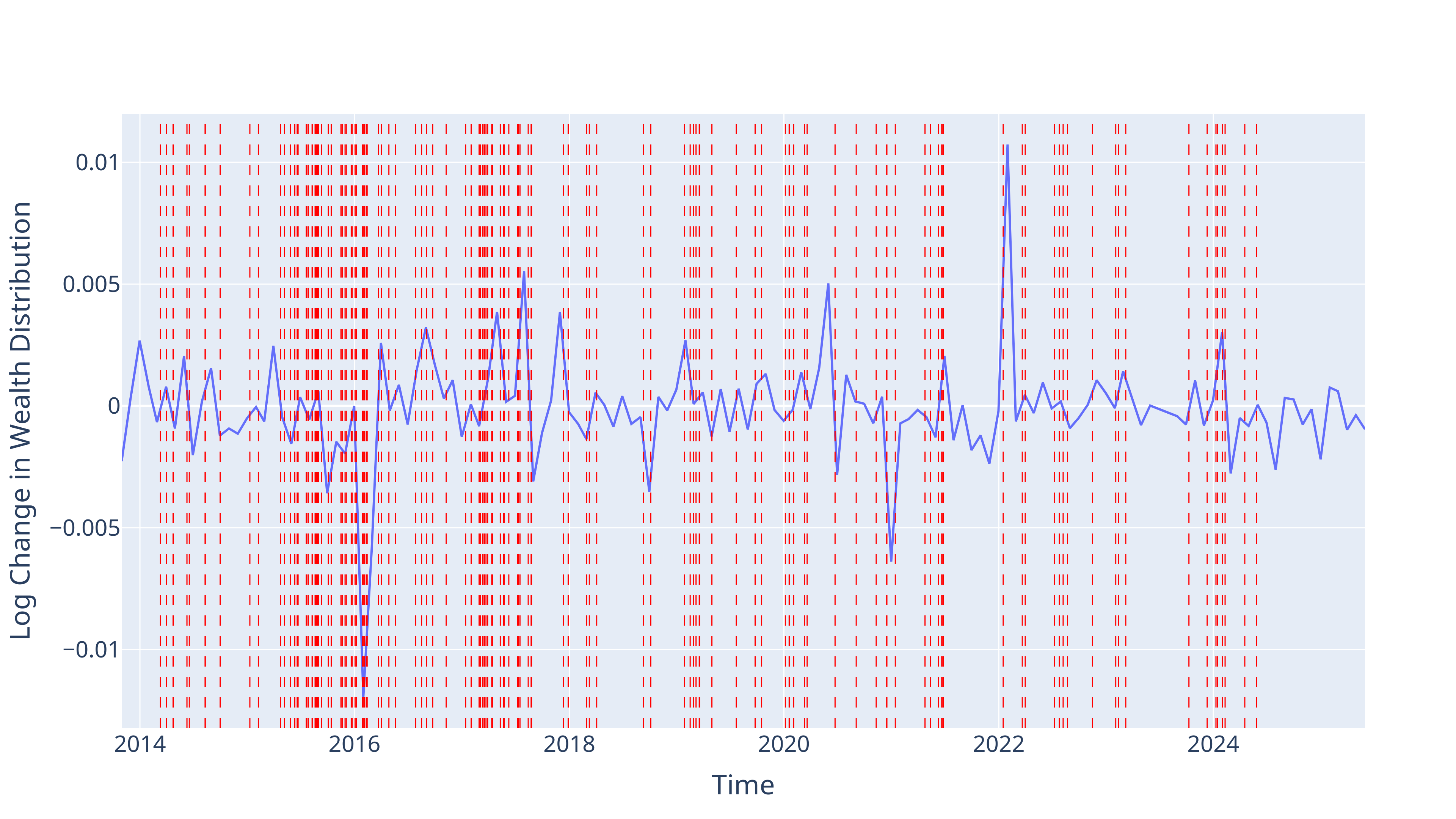}
	\caption{Log change in wealth distribution of bucket \emph{From 100 to 1000}}
	\includegraphics[width=1\linewidth]{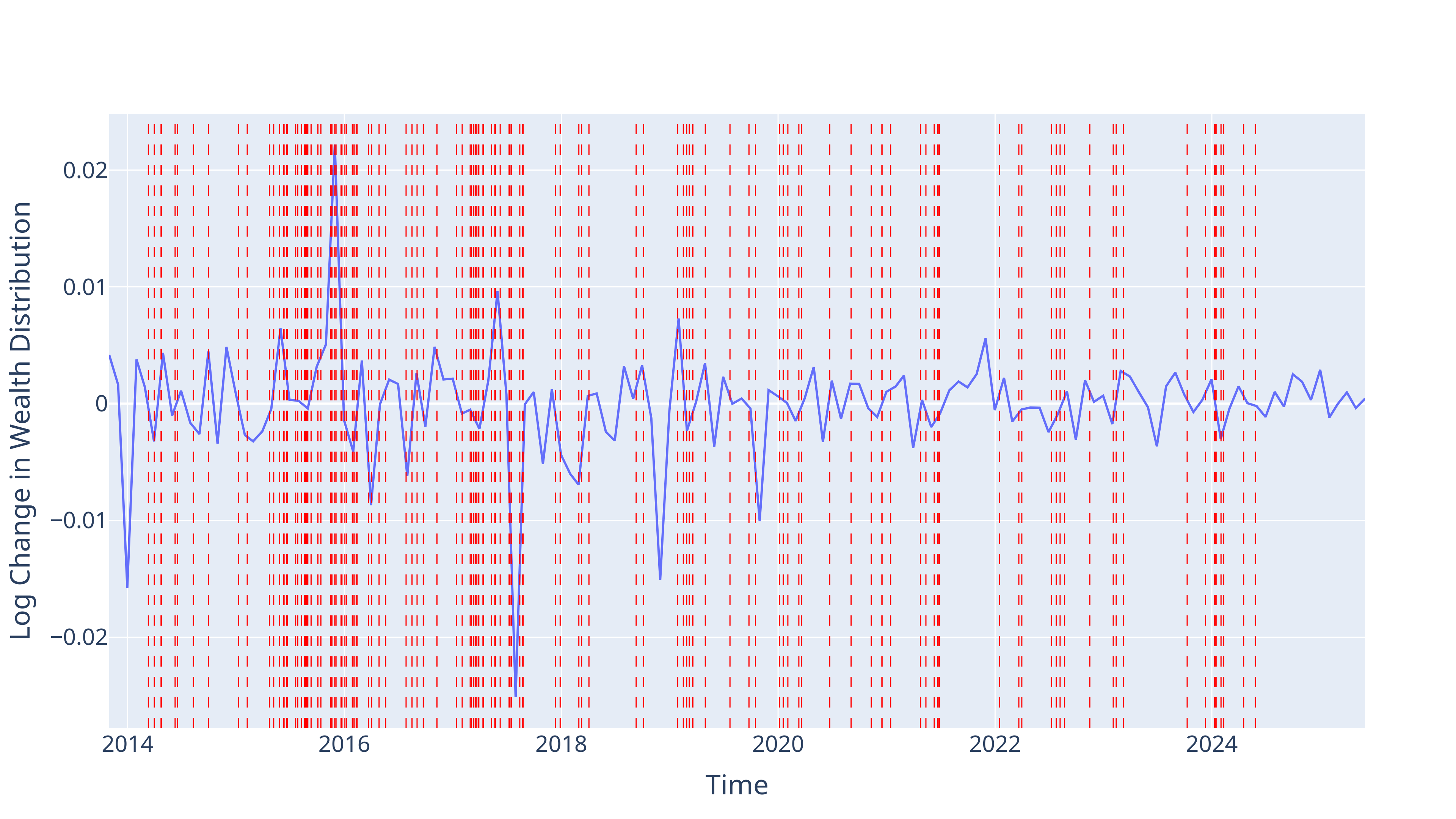}
	\caption{Log change in wealth distribution of bucket \emph{From 1000 to 10000}}
\end{figure}
\begin{figure}[ht!]
	\centering
	\includegraphics[width=1\linewidth]{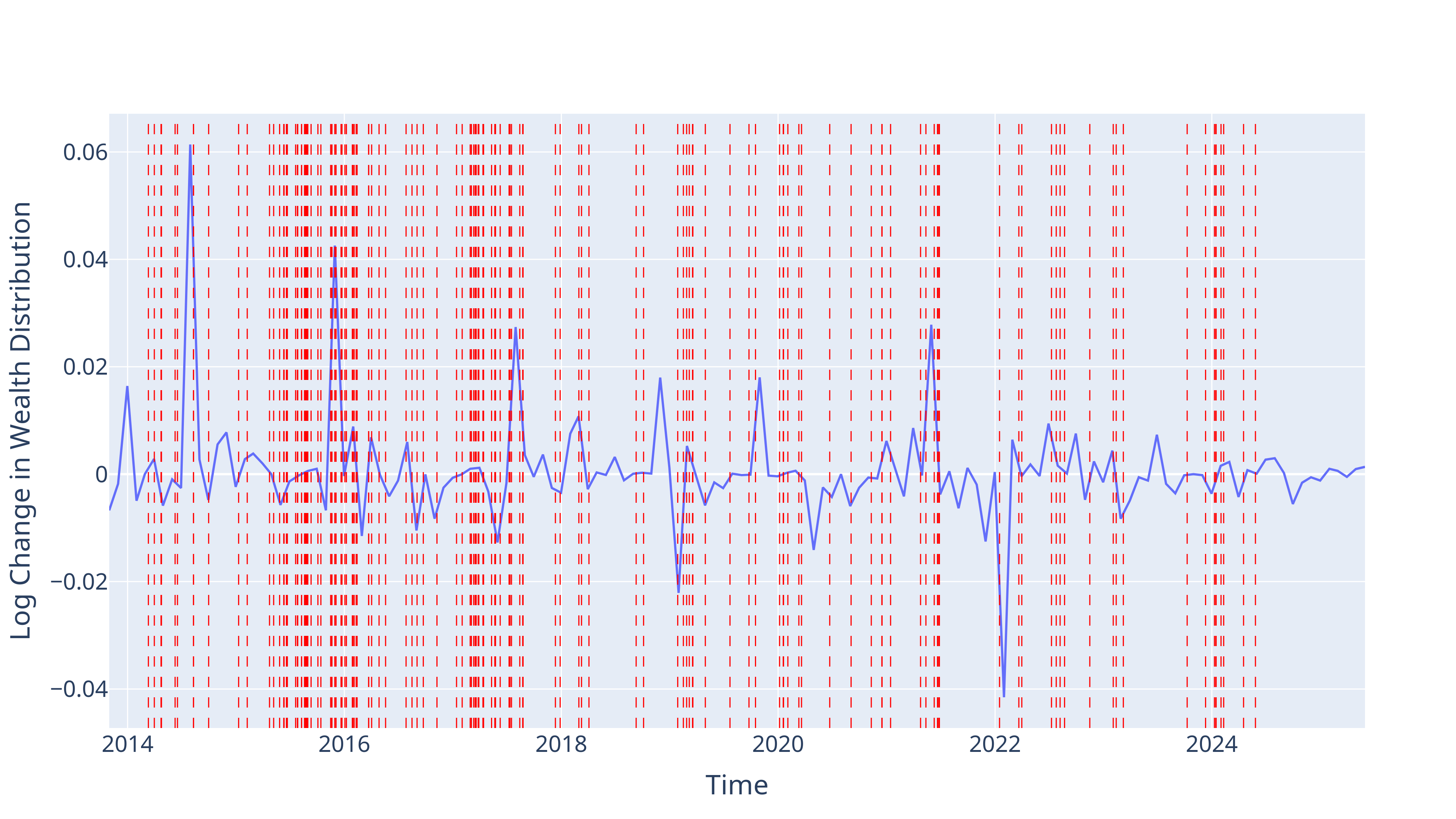}
	\caption{Log change in wealth distribution of bucket \emph{From 10000 to 100000}}
	\includegraphics[width=1\linewidth]{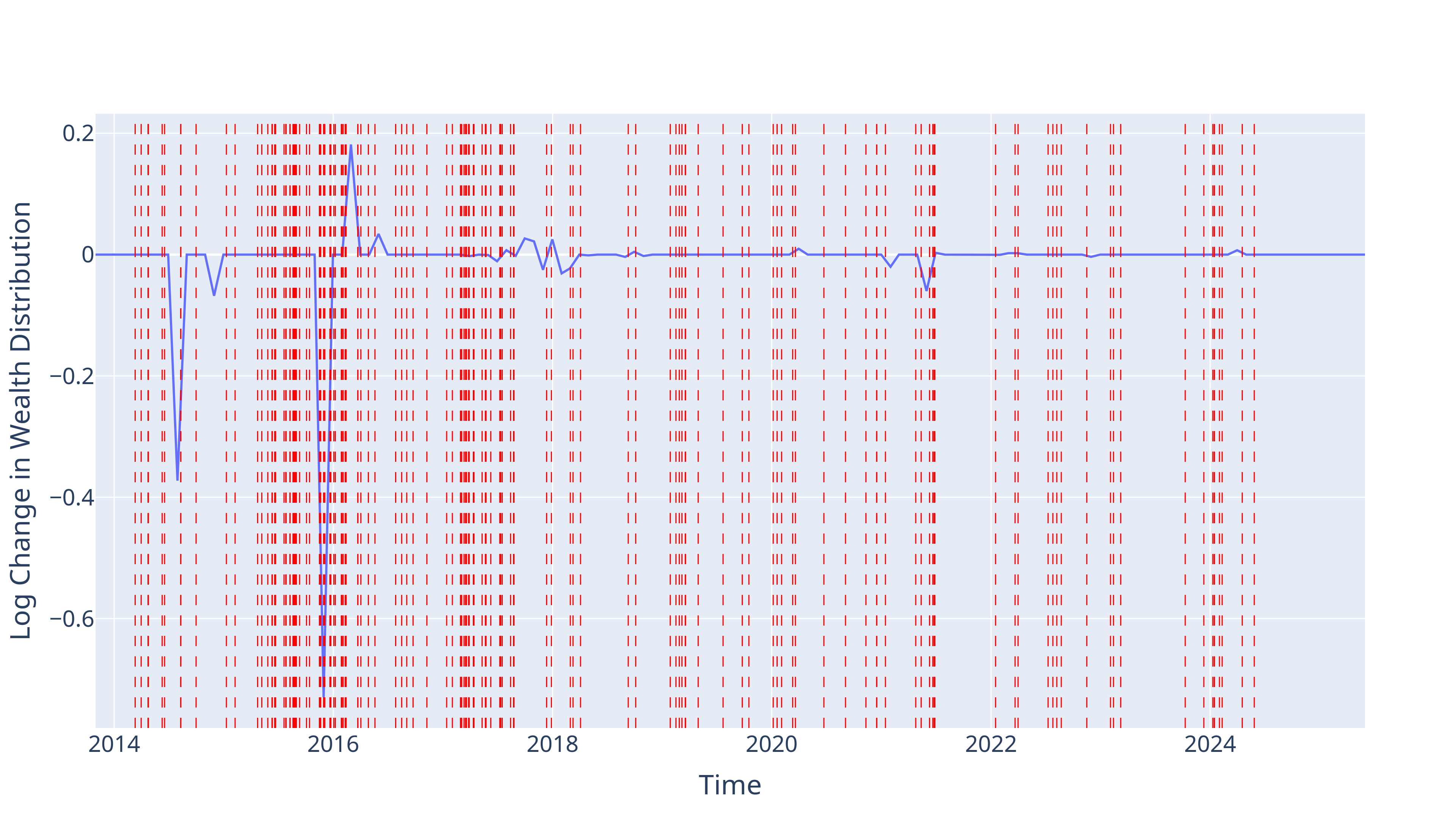}
	\caption{Log change in wealth distribution of bucket \emph{From 100000 to infinity}}
\end{figure}

\clearpage
\newpage
\section{Global Linear Regression (p-values)}
\label{appendix:GlobalLinearRegressionPValues}
\textbf{Hint}: Asterisk and cell shading highlight the p-values below the chosen critical level of $5\%$.
\begin{center}
\begin{tabular}{ |c|c|c|c|c|c| }
\hline
 & \rotatebox{90}{From 0 to 0.001} & \rotatebox{90}{From 0.001 to 0.01 } & \rotatebox{90}{From 0.01 to 0.1} & \rotatebox{90}{From 0.1 to 1} & \rotatebox{90}{From 1 to 10} \\ 
\hline
Bitcoin (Close Price) & 0.95552 & 0.05879 & 0.11609 & 0.39121 & 0.2371 \\ 
\hline
Gold Price Against USD & 0.78693 & 0.77605 & 0.17878 & 0.24755 & 0.75837 \\ 
\hline
\begin{tabular}{@{}c@{}}20 Year GILTs \\ (Nominal Zero Coupon Yield)\end{tabular} & \cellcolor{green!25} $0.03687^{\ast}$ & 0.33783 & 0.8859 & 0.67275 & 0.73541 \\ 
\hline
\begin{tabular}{@{}c@{}}10 Year GILTs \\ (Nominal Zero Coupon Yield)\end{tabular} & 0.83118 & 0.30361 & 0.10171 & 0.14961 & 0.3184 \\ 
\hline
\begin{tabular}{@{}c@{}}5 Year GILTs \\ (Nominal Zero Coupon Yield)\end{tabular} & 0.30856 & 0.96274 & 0.78562 & 0.79379 & 0.89922 \\ 
\hline
\begin{tabular}{@{}c@{}}20 Year GILTs \\ (Nominal Par Yield)\end{tabular} & 0.68882 & 0.84201 & 0.80301 & 0.71128 & 0.49422 \\ 
\hline
\begin{tabular}{@{}c@{}}10 Year GILTs \\ (Nominal Par Yield)\end{tabular} & 0.47746 & 0.65285 & 0.33442 & 0.58821 & 0.51561 \\ 
\hline
\begin{tabular}{@{}c@{}}5 Year GILTs \\ (Nominal Par Yield)\end{tabular} & 0.57838 & \cellcolor{green!25} $0.0172^{\ast}$ & 0.48107 & 0.24873 & 0.07603 \\ 
\hline
\begin{tabular}{@{}c@{}c@{}}Producer Price Index \\ (Semiconductors and \\ Electronic Components)\end{tabular} & 0.59451 & 0.97032 & 0.57565 & 0.59474 & 0.71034 \\ 
\hline
\begin{tabular}{@{}c@{}}30 Year High Quality Market \\ Corporate Bond Spot Rate\end{tabular} & 0.90286 & 0.75138 & 0.41004 & 0.24977 & 0.70001 \\ 
\hline
\begin{tabular}{@{}c@{}}10 Year High Quality Market \\ Corporate Bond Spot Rate\end{tabular} & 0.75744 & 0.50603 & 0.79627 & 0.31648 & 0.85344 \\ 
\hline
\begin{tabular}{@{}c@{}}5 Year High Quality Market \\ Corporate Bond Spot Rate\end{tabular} & 0.12468 & 0.4359 & 0.36901 & 0.74853 & 0.44947 \\ 
\hline
\begin{tabular}{@{}c@{}}30 Year High Quality Market \\ Corporate Bond Par Yield\end{tabular} & 0.0981 & 0.35661 & 0.55473 & 0.73128 & 0.33157 \\ 
\hline
\begin{tabular}{@{}c@{}}10 Year High Quality Market \\ Corporate Bond Par Yield\end{tabular} & 0.34237 & 0.4497 & 0.25276 & 0.67013 & 0.16626 \\ 
\hline
\begin{tabular}{@{}c@{}}5 Year High Quality Market \\ Corporate Bond Par Yield\end{tabular} & 0.88041 & 0.58471 & 0.61756 & 0.21268 & 0.59216 \\ 
\hline
Unemployment Rate (US) & 0.83498 & 0.07568 & 0.72303 & 0.76021 & 0.06659 \\ 
\hline
Consumer Price Index (US) & 0.65166 & 0.89672 & 0.72195 & 0.88553 & 0.91983 \\ 
\hline
\begin{tabular}{@{}c@{}}Market Value of Privately Held \\ Gross Federal Debt\end{tabular} & 0.96968 & 0.81238 & 0.91815 & 0.70745 & 0.94792 \\ 
\hline
\begin{tabular}{@{}c@{}}Market Value of \\ Gross Federal Debt\end{tabular} & 0.11826 & 0.23086 & 0.10521 & 0.39906 & 0.10054 \\ 
\hline
M2 (US) & 0.79448 & 0.24604 & 0.46469 & 0.20304 & 0.86395 \\ 
\hline
M1 (US) & 0.99723 & 0.28262 & 0.32516 & 0.11052 & 0.33234 \\ 
\hline
Nonfarm Payroll (US) & 0.84165 & \cellcolor{green!25} $0.04645^{\ast}$ & 0.70288 & 0.73462 & \cellcolor{green!25} $0.04578^{\ast}$ \\ 
\hline
Personal Current Taxes (US) & 0.06454 & 0.12047 & 0.36148 & 0.52582 & 0.57357 \\ 
\hline
Federal Funds Rate & \cellcolor{green!25} $0.01864^{\ast}$ & 0.91238 & 0.80898 & 0.98497 & 0.10087 \\ 
\hline
\end{tabular}
\end{center}

\begin{center}
\begin{tabular}{ |c|c|c|c|c|c| }
\hline
 & \rotatebox{90}{From 10 to 100} & \rotatebox{90}{From 100 to 1000} & \rotatebox{90}{From 1000 to 10000} & \rotatebox{90}{From 10000 to 100000} & \rotatebox{90}{From 100000 to infinity } \\ 
\hline
Bitcoin (Close Price) & 0.93867 & 0.46682 & 0.70236 & 0.21032 & 0.80568 \\ 
\hline
Gold Price Against USD & 0.76503 & 0.68464 & 0.07651 & 0.6149 & 0.68218 \\ 
\hline
\begin{tabular}{@{}c@{}}20 Year GILTs \\ (Nominal Zero Coupon Yield)\end{tabular} & 0.51814 & 0.43059 & 0.80933 & 0.34878 & 0.40557 \\ 
\hline
\begin{tabular}{@{}c@{}}10 Year GILTs \\ (Nominal Zero Coupon Yield)\end{tabular} & 0.89177 & 0.69377 & 0.2441 & 0.28757 & 0.90289 \\ 
\hline
\begin{tabular}{@{}c@{}}5 Year GILTs \\ (Nominal Zero Coupon Yield)\end{tabular} & 0.35175 & 0.68574 & 0.38014 & 0.59197 & 0.22758 \\ 
\hline
\begin{tabular}{@{}c@{}}20 Year GILTs \\ (Nominal Par Yield)\end{tabular} & \cellcolor{green!25} $0.01138^{\ast}$ & 0.30415 & 0.25454 & 0.82128 & 0.39393 \\ 
\hline
\begin{tabular}{@{}c@{}}10 Year GILTs \\ (Nominal Par Yield)\end{tabular} & 0.64698 & 0.46399 & 0.63107 & 0.59572 & 0.19319 \\ 
\hline
\begin{tabular}{@{}c@{}}5 Year GILTs \\ (Nominal Par Yield)\end{tabular} & 0.4147 & 0.17991 & 0.80825 & 0.44746 & 0.64333 \\ 
\hline
\begin{tabular}{@{}c@{}c@{}}Producer Price Index \\ (Semiconductors and \\ Electronic Components)\end{tabular} & 0.63546 & 0.19781 & 0.8154 & 0.42572 & 0.37454 \\ 
\hline
\begin{tabular}{@{}c@{}}30 Year High Quality Market \\ Corporate Bond Spot Rate\end{tabular} & 0.57632 & 0.66095 & 0.43108 & 0.54761 & 0.36277 \\ 
\hline
\begin{tabular}{@{}c@{}}10 Year High Quality Market \\ Corporate Bond Spot Rate\end{tabular} & 0.41708 & 0.56404 & 0.72565 & 0.26456 & 0.52256 \\ 
\hline
\begin{tabular}{@{}c@{}}5 Year High Quality Market \\ Corporate Bond Spot Rate\end{tabular} & 0.16551 & 0.39963 & 0.96729 & 0.12264 & 0.52284 \\ 
\hline
\begin{tabular}{@{}c@{}}30 Year High Quality Market \\ Corporate Bond Par Yield\end{tabular} & 0.37634 & 0.39991 & 0.51794 & 0.47668 & 0.62582 \\ 
\hline
\begin{tabular}{@{}c@{}}10 Year High Quality Market \\ Corporate Bond Par Yield\end{tabular} & 0.62523 & 0.38101 & 0.28999 & 0.74516 & 0.42881 \\ 
\hline
\begin{tabular}{@{}c@{}}5 Year High Quality Market \\ Corporate Bond Par Yield\end{tabular} & 0.83063 & 0.65482 & \cellcolor{green!25} $0.03731^{\ast}$ & 0.85632 & 0.50252 \\ 
\hline
Unemployment Rate (US) & 0.52021 & 0.58443 & 0.64602 & 0.09818 & 0.92865 \\ 
\hline
Consumer Price Index (US) & 0.20949 & 0.89747 & \cellcolor{green!25} $0.02143^{\ast}$ & 0.43401 & 0.63789 \\ 
\hline
\begin{tabular}{@{}c@{}}Market Value of Privately Held \\ Gross Federal Debt\end{tabular} & 0.33362 & 0.16659 & 0.06504 & 0.55106 & 0.95536 \\ 
\hline
\begin{tabular}{@{}c@{}}Market Value of \\ Gross Federal Debt\end{tabular} & 0.57143 & 0.61409 & 0.95303 & 0.59789 & \cellcolor{green!25} $0.01417^{\ast}$ \\ 
\hline
M2 (US) & \cellcolor{green!25} $0.02234^{\ast}$ & 0.63307 & 0.89056 & 0.16755 & 0.57691 \\ 
\hline
M1 (US) & 0.91529 & 0.43484 & 0.97615 & 0.07317 & 0.79659 \\ 
\hline
Nonfarm Payroll (US) & 0.52601 & 0.4809 & 0.80202 & 0.12235 & 0.84016 \\ 
\hline
Personal Current Taxes (US) & 0.72656 & 0.35213 & 0.12155 & 0.30312 & 0.88685 \\ 
\hline
Federal Funds Rate & 0.06506 & 0.30516 & 0.21317 & 0.87413 & 0.16522 \\ 
\hline
\end{tabular}
\end{center}

\clearpage
\newpage
\section{BIPs Causality Analysis}
\label{appendix:BIPsCausalityAnalysis}
\textbf{True}: Accept the assumption of Granger-causality
\begin{itemize}
	\item \textbf{T (x)}: $x$ is the longest significant lag (in months) of the independent variable in the parent model in the Granger-causality test.
\end{itemize}
\textbf{False}: Reject the assumption of Granger-causality
\begin{itemize}
\setlength\itemsep{-0.5em}
	\item \textbf{F (a)}: Could not select a model with autoregressive terms of the independent variable as all models have infinite Akaike information criterion.
	\item \textbf{F (b)}: All autoregressive terms of the independent variable have been filtered out as insignificant according to the t-test.
	\item \textbf{F (c)}: Autoregressive terms of the independent variable do not add explanatory power according to the F-test.
\end{itemize}
\begin{center}
\begin{tabular}{ |c|c|c|c|c|c|c|c|c| }
\hline
 & \rotatebox{90}{Simple Test, All BIPs} & \rotatebox{90}{Simple Test, All Economy-Related BIPs} & \rotatebox{90}{Simple Test, Major Economy-Related BIPs} & \rotatebox{90}{Simple Test, All Economy-Related BIPs (Except the major ones) } & \rotatebox{90}{Full Test, All BIPs} & \rotatebox{90}{Full Test, All Economy-Related BIPs} & \rotatebox{90}{Full Test, Major Economy-Related BIPs} & \rotatebox{90}{Full Test, All Economy-Related BIPs (Except the major ones)} \\ 
\hline
From 0 to 0.001 & F (c) & F (c) & F (c) & F (c) & F (b) & F (b) & F (b) & F (b) \\ 
\hline
From 0.001 to 0.01 & F (c) & F (c) & \cellcolor{green!25} T (10) & F (c) & F (b) & F (b) & \cellcolor{green!25} T (3) & F (b) \\ 
\hline
From 0.01 to 0.1 & F (c) & F (c) & \cellcolor{green!25} T (10) & F (c) & F (b) & F (b) & \cellcolor{green!25} T (10) & F (b) \\ 
\hline
From 0.1 to 1 & F (c) & F (c) & \cellcolor{green!25} T (10) & F (c) & F (b) & F (b) & \cellcolor{green!25} T (1) & F (b) \\ 
\hline
From 1 to 10 & F (c) & F (c) & \cellcolor{green!25} T (10) & F (c) & F (b) & F (b) & F (b) & F (b) \\ 
\hline
From 10 to 100 & F (c) & F (c) & F (c) & \cellcolor{green!25} T (10) & F (b) & F (b) & F (b) & F (b) \\ 
\hline
From 100 to 1000 & F (c) & F (c) & \cellcolor{green!25} T (10) & F (c) & F (b) & F (b) & F (c) & F (b) \\ 
\hline
From 1000 to 10000 & F (c) & F (c) & F (c) & F (c) & F (b) & \cellcolor{green!25} T (3) & F (b) & \cellcolor{green!25} T (3) \\ 
\hline
From 10000 to 100000 & F (c) & \cellcolor{green!25} T (10) & F (c) & \cellcolor{green!25} T (10) & F (b) & \cellcolor{green!25} T (6) & F (b) & \cellcolor{green!25} T (6) \\ 
\hline
From 100000 to infinity & F (c) & F (c) & F (c) & F (c) & F (b) & F (b) & F (b) & F (b) \\ 
\hline
\end{tabular}
\end{center}

\clearpage
\newpage
\section{BIPs Causality Analysis (6 months)}
\label{appendix:BIPsCausalityAnalysis6Months}
\textbf{True}: Accept the assumption of Granger-causality
\begin{itemize}
	\item \textbf{T (x)}: $x$ is the longest significant lag (in months) of the independent variable in the parent model in the Granger-causality test.
\end{itemize}
\textbf{False}: Reject the assumption of Granger-causality
\begin{itemize}
\setlength\itemsep{-0.5em}
	\item \textbf{F (a)}: Could not select a model with autoregressive terms of the independent variable as all models have infinite Akaike information criterion.
	\item \textbf{F (b)}: All autoregressive terms of the independent variable have been filtered out as insignificant according to the t-test.
	\item \textbf{F (c)}: Autoregressive terms of the independent variable do not add explanatory power according to the F-test.
\end{itemize}
\begin{tabular}{ |c|c|c|c|c|c|c|c|c| }
\hline
 & \rotatebox{90}{Simple Test, All BIPs} & \rotatebox{90}{Simple Test, All Economy-Related BIPs} & \rotatebox{90}{Simple Test, Major Economy-Related BIPs} & \rotatebox{90}{Simple Test, All Economy-Related BIPs (Except the major ones)} & \rotatebox{90}{Full Test, All BIPs} & \rotatebox{90}{Full Test, All Economy-Related BIPs} & \rotatebox{90}{Full Test, Major Economy-Related BIPs} & \rotatebox{90}{Full Test, All Economy-Related BIPs (Except the major ones)} \\ 
\hline
From 0 to 0.001 & F (c) & F (c) & F (c) & F (c) & F (b) & F (b) & F (b) & F (b) \\ 
\hline
From 0.001 to 0.01 & F (c) & F (c) & \cellcolor{green!25} T (6) & F (c) & F (b) & F (b) & \cellcolor{green!25} T (3) & F (b) \\ 
\hline
From 0.01 to 0.1 & F (c) & F (c) & F (c) & F (c) & F (b) & F (b) & \cellcolor{green!25} T (6) & F (b) \\ 
\hline
From 0.1 to 1 & F (c) & F (c) & \cellcolor{green!25} T (6) & F (c) & F (b) & F (b) & \cellcolor{green!25} T (1) & F (b) \\ 
\hline
From 1 to 10 & F (c) & F (c) & F (c) & F (c) & F (b) & F (b) & F (b) & F (b) \\ 
\hline
From 10 to 100 & F (c) & \cellcolor{green!25} T (6) & F (c) & \cellcolor{green!25} T (6) & F (b) & F (b) & F (b) & F (b) \\ 
\hline
From 100 to 1000 & F (c) & F (c) & \cellcolor{green!25} T (6) & F (c) & F (b) & F (b) & F (c) & F (b) \\ 
\hline
From 1000 to 10000 & F (c) & F (c) & F (c) & F (c) & F (b) & \cellcolor{green!25} T (3) & F (b) & \cellcolor{green!25} T (3) \\ 
\hline
From 10000 to 100000 & F (c) & F (c) & F (c) & F (c) & F (c) & \cellcolor{green!25} T (1) & \cellcolor{green!25} T (3) & \cellcolor{green!25} T (1) \\ 
\hline
From 100000 to infinity & F (c) & F (c) & F (c) & F (c) & F (b) & F (b) & F (b) & F (b) \\ 
\hline
\end{tabular}

\clearpage
\newpage
\section{BIPs Causality Analysis (12 months)}
\label{appendix:BIPsCausalityAnalysis12Months}
\textbf{True}: Accept the assumption of Granger-causality
\begin{itemize}
	\item \textbf{T (x)}: $x$ is the longest significant lag (in months) of the independent variable in the parent model in the Granger-causality test.
\end{itemize}
\textbf{False}: Reject the assumption of Granger-causality
\begin{itemize}
\setlength\itemsep{-0.5em}
	\item \textbf{F (a)}: Could not select a model with autoregressive terms of the independent variable as all models have infinite Akaike information criterion.
	\item \textbf{F (b)}: All autoregressive terms of the independent variable have been filtered out as insignificant according to the t-test.
	\item \textbf{F (c)}: Autoregressive terms of the independent variable do not add explanatory power according to the F-test.
\end{itemize}
\begin{tabular}{ |c|c|c|c|c|c|c|c|c| }
\hline
 & \rotatebox{90}{Simple Test, All BIPs} & \rotatebox{90}{Simple Test, All Economy-Related BIPs} & \rotatebox{90}{Simple Test, Major Economy-Related BIPs} & \rotatebox{90}{Simple Test, All Economy-Related BIPs (Except the major ones)} & \rotatebox{90}{Full Test, All BIPs} & \rotatebox{90}{Full Test, All Economy-Related BIPs} & \rotatebox{90}{Full Test, Major Economy-Related BIPs} & \rotatebox{90}{Full Test, All Economy-Related BIPs (Except the major ones)} \\ 
\hline
From 0 to 0.001 & F (c) & F (c) & F (c) & F (c) & F (b) & F (b) & F (b) & F (c) \\ 
\hline
From 0.001 to 0.01 & F (c) & F (c) & \cellcolor{green!25} T (12) & F (c) & F (b) & F (b) & \cellcolor{green!25} T (3) & F (b) \\ 
\hline
From 0.01 to 0.1 & F (c) & F (c) & F (c) & F (c) & F (b) & F (b) & F (b) & F (b) \\ 
\hline
From 0.1 to 1 & F (c) & F (c) & \cellcolor{green!25} T (12) & F (c) & F (b) & F (b) & \cellcolor{green!25} T (1) & F (b) \\ 
\hline
From 1 to 10 & F (c) & F (c) & F (c) & F (c) & F (b) & F (b) & F (b) & F (b) \\ 
\hline
From 10 to 100 & F (c) & F (c) & F (c) & \cellcolor{green!25} T (12) & F (b) & F (b) & F (b) & F (b) \\ 
\hline
From 100 to 1000 & F (c) & F (c) & F (c) & F (c) & F (b) & F (b) & F (c) & F (b) \\ 
\hline
From 1000 to 10000 & F (c) & F (c) & F (c) & F (c) & F (b) & \cellcolor{green!25} T (3) & F (b) & \cellcolor{green!25} T (3) \\ 
\hline
From 10000 to 100000 & F (c) & \cellcolor{green!25} T (12) & F (c) & \cellcolor{green!25} T (12) & F (c) & \cellcolor{green!25} T (6) & F (b) & \cellcolor{green!25} T (6) \\ 
\hline
From 100000 to infinity & F (c) & F (c) & F (c) & F (c) & F (b) & F (b) & F (b) & F (b) \\ 
\hline
\end{tabular}

\clearpage
\newpage
\section{Taxonomised BIPs Causality Analysis}
\label{appendix:TaxonomisedBIPsCausalityAnalysis}
\textbf{True}: Accept the assumption of Granger-causality
\begin{itemize}
	\item \textbf{T (x)}: $x$ is the longest significant lag (in months) of the independent variable in the parent model in the Granger-causality test.
\end{itemize}
\textbf{False}: Reject the assumption of Granger-causality
\begin{itemize}
\setlength\itemsep{-0.5em}
	\item \textbf{F (a)}: Could not select a model with autoregressive terms of the independent variable as all models have infinite Akaike information criterion.
	\item \textbf{F (b)}: All autoregressive terms of the independent variable have been filtered out as insignificant according to the t-test.
	\item \textbf{F (c)}: Autoregressive terms of the independent variable do not add explanatory power according to the F-test.
\end{itemize}
\begin{center}
\begin{tabular}{ |c|c|c|c|c|c|c| }
\hline
 & \rotatebox{90}{Simple Test, Fiscal-Like BIPs} & \rotatebox{90}{Simple Test, Monetary-Like BIPs} & \rotatebox{90}{Simple Test, Purely Tokenomic BIPs } & \rotatebox{90}{Full Test, Fiscal-Like BIPs} & \rotatebox{90}{Full Test, Monetary-Like BIPs} & \rotatebox{90}{Full Test, Purely Tokenomic BIPs} \\ 
\hline
From 0 to 0.001 & F (c) & F (c) & F (c) & F (b) & F (b) & F (b) \\ 
\hline
From 0.001 to 0.01 & F (c) & \cellcolor{green!25} T (10) & \cellcolor{green!25} T (10) & F (b) & \cellcolor{green!25} T (10) & F (b) \\ 
\hline
From 0.01 to 0.1 & F (c) & F (c) & F (c) & \cellcolor{green!25} T (9) & F (b) & F (b) \\ 
\hline
From 0.1 to 1 & F (c) & \cellcolor{green!25} T (10) & F (c) & F (b) & \cellcolor{green!25} T (8) & \cellcolor{green!25} T (5) \\ 
\hline
From 1 to 10 & \cellcolor{green!25} T (10) & F (c) & \cellcolor{green!25} T (10) & \cellcolor{green!25} T (4) & F (b) & \cellcolor{green!25} T (2) \\ 
\hline
From 10 to 100 & F (c) & \cellcolor{green!25} T (10) & F (c) & F (b) & F (b) & F (b) \\ 
\hline
From 100 to 1000 & F (c) & F (c) & \cellcolor{green!25} T (10) & F (b) & F (b) & \cellcolor{green!25} T (7) \\ 
\hline
From 1000 to 10000 & \cellcolor{green!25} T (10) & F (c) & F (c) & F (b) & F (b) & \cellcolor{green!25} T (3) \\ 
\hline
From 10000 to 100000 & F (c) & F (c) & F (c) & \cellcolor{green!25} T (5) & \cellcolor{green!25} T (6) & \cellcolor{green!25} T (1) \\ 
\hline
From 100000 to infinity & F (c) & \cellcolor{green!25} T (10) & F (c) & F (b) & F (b) & F (b) \\ 
\hline
\end{tabular}
\end{center}

\end{document}